\documentclass[onecolumn]{emulateapj}

\usepackage{amssymb}
\usepackage{wrapfig}
\usepackage{enumitem}

\begin{document}

\title{Astrochemistry and compositions of planetary systems}

\author{Karin I. \"Oberg}
\affil{Harvard-Smithsonian Center for Astrophysics, 60 Garden St., Cambridge, MA 02138}


\author{Edwin A. Bergin}
\affil{Department of Astronomy, University of Michigan,
1085 S. University Avenue,
Ann Arbor, MI 48109, USA
}

\address{}

\begin{abstract}
Planets form and obtain their compositions in disks of gas and dust around young stars. The chemical compositions of these planet-forming disks regulate all aspects of planetary compositions from bulk elemental inventories to access to water and reactive organics, i.e. a planet's hospitality to life and its chemical origins. Disk chemical structures are in their turn governed by a combination of {\it in situ} chemical processes, and inheritance of molecules from the preceding evolutionary stages of the star formation process. In this review we present our current understanding of the chemical processes active in pre- and protostellar environments that set the initial conditions for disks, and the disk chemical processes that evolve the chemical conditions during the first million years of planet formation. We review recent observational, laboratory and theoretical discoveries that have led to the present view of the chemical environment within which planets form, and their effects on the compositions of nascent planetary systems. We also discuss the many unknowns that remain and outline some possible pathways to addressing them.
\end{abstract}

\section{Introduction}

The goal of this review is to present the astrochemistry most relevant to predicting and interpreting the volatile compositions of planetary systems. The past decades have witnessed a revolution in planetary science, increasing the number of known planets from eight around the Sun to many thousands of exoplanets around other stars. Many of these exoplanets are strange compared to expectations from the solar system \citep{Winn15,Dawson18}, and planetary compositions are emerging as a major tool to constrain the origins of specific planets, as well as families of planets \citep{Oberg11d,Madhusudhan19}. Doing this well requires a detailed understanding of the elemental composition of the gas and solids from which planets form. Furthermore, a subset of the known exoplanets are Earth-like in the sense that they are rocky and temperate, i.e. at the right temperature to sustain liquid water. 
For this set of potentially habitable exoplanets, we have a second set of compositional questions which relates to their hospitality for life and origins of life chemistry: do they form with water and ready access to reactive organic molecules, sulfur and phosphorus? These questions too depend on the composition of the planet-forming material, but this time on the detailed molecular composition.

Planets form in disks of gas and dust around young stars \citep[e.g.][]{Lissauer93,Andrews18}. The complete composition of dust and gas is relevant to predicting planet compositions, but in this review we will mostly limit ourselves to the volatile components, i.e. the gas and icy grain mantles. Based on theory as well as on direct observations of disks, and data from our own solar system, disk compositions are in their turn a product of both {\it in situ} disk chemical processes, and inheritance of molecular material from the preceding evolutionary stages \citep{Mumma11,Cleeves14,Ceccarelli14,Drozdovskaya14,Huang17}. A complete model of the astrochemical origins of compositions of planetary systems must then include the chemical evolution from the onset of cloud formation, via star formation within molecular clouds, to the final stages of planet formation in disks.

In the remainder of this section we briefly review the formation of Solar-type stars and their accompanying planetary systems, astrochemical reactions, and astrochemical methods. This is followed by a section on the chemical foundations, i.e. the division of elements between volatile carriers, and the formation of water and of the first organics, that develop in molecular clouds and cloud cores. In the subsequent section we review how the chemistry evolves in the protostellar stage, and existing constraints on the chemical composition of the protostellar precursors to the mature, planet-forming disks. The final section treats observations and theory of chemistry in planet-forming disks and how they relate to observed volatile compositions within the solar system.

\subsection{Formation of stars and planetary systems}

\begin{figure}[htbp]
\begin{center}
\includegraphics[width=0.8\textwidth]{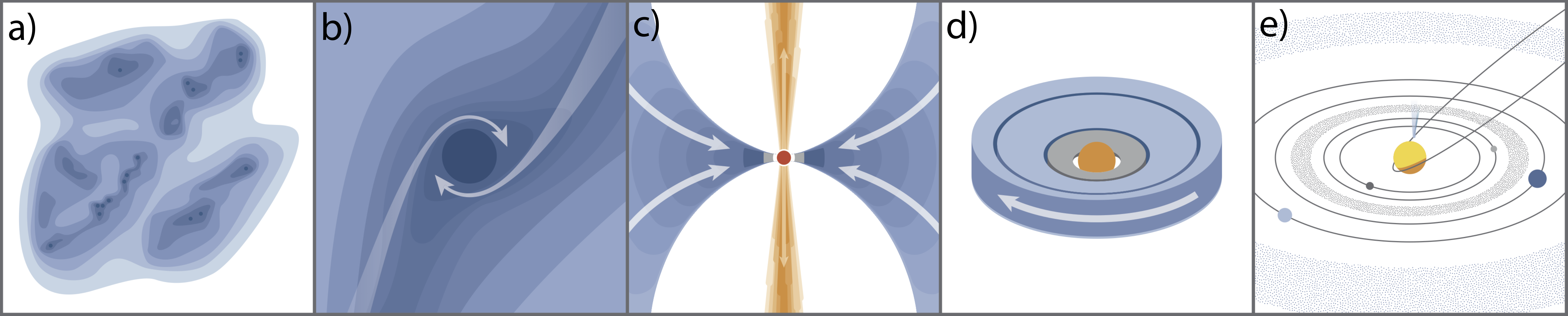}
\caption{Cartoon of the different stages characterizing low-mass (Solar-like) star and planet formation. {\bf a:} Stars form in dense cores in interstellar molecular clouds. {\bf b:} Star formation begins when such a dense core begins to collapse due to self-gravity. {\bf c}: As the collapse proceeds the center heats up forming a protostar. Accretion of remnant cloud material continues, funneled through a disk, which is formed as a consequence of cloud angular momentum. This stage is also characterized by outflows of material. {\bf d:} Following dispersal of the cloud remnant the now pre-main sequence star is left with a circumstellar disk, which is the formation sites of planets. The disk gas is dispersed through disk winds within $\sim$2-5 Myrs, putting a halt to Gas Giant formation. {\bf e:} Rocky and icy planets can continue to grow for another 100 Myrs at which time a mature planetary system exists. Image credit: K. Peek.}
\label{fig:shu}
\end{center}
\end{figure}

Stars span a large range of masses, and the star formation process depends on the mass of the final star. For the purpose of predicting planetary compositions, low-mass stars (less than two solar masses) are the most relevant. They constitute the vast majority of stars and therefore planet hosts in the Galaxy, and their long life times are likely a pre-requisite for the origins of life. 
There are several excellent reviews of low-mass star formation going back to \cite{Shu87}, whose illustration of low-mass star formation we have updated in Figure \ref{fig:shu}. More recently star formation has been reviewed by \cite{McKee07} and \cite{Luhman12}. Here we present a brief summary of the current understanding of low-mass star formation, which will be the context for the remainder of the review.

Stars form in over-dense regions in the interstellar medium, referred to as clouds (Fig. \ref{fig:shu}a). Clouds mostly consist of gas, but approximately one percent of the mass of the cloud is in the form of sub-micron dust particles, initially composed of silicates and carbonaceous material \citep{Draine03_araa, Henning10, Chiar13, Jones13}. These dust grains can become coated with icy mantles of more volatile species, which have a profound impact on the chemical evolution during star formation \citep{Herbst09}. Star formation is mainly associated with dense molecular clouds \citep{Heyer15_araa}, where densities are $>10^2$ $n_{\rm H}$ cm$^{-3}$ (hydrogen nuclei per cubic centimeter), which result in cloud interiors that are therefore well shielded from external radiation, resulting in low temperature and most elements being bound up in molecules. UV radiation chemistry is still key to understand cloud chemical structures, however, since competition between molecular formation and photodissociation during cloud assembly determines the initial chemical compositions of dense clouds \citep{Bergin04, Glover07, Clark12, Seifried17}.

Figure \ref{fig:shu}a also illustrates that within dense clouds there are even denser sub-structures referred to as dense cores. Compared to the large scale dense cloud, these are characterized by orders of magnitude higher densities, $\sim$10$^{6}$ vs. $\sim$10$^{2}-10^4$, and lower temperatures, on the order of 10~K \citep{Benson89,Bergin07}. Typical core sizes are on the order of a tenth of a parsec.  Some dense cores are dense enough that they can begin to collapse, overcoming turbulence, thermal and magnetic pressure. Cores that are fated to collapse are referred to as pre-stellar. This collapse is initially isothermal and somewhat asymmetrical due to preservation of cloud angular momentum (Fig. \ref{fig:shu}b). As the collapse proceeds, the center of the collapsing core becomes optically thick and heats up: a protostar has formed (Fig. \ref{fig:shu}c). 

During the protostellar stage the central protostar continues to accrete mass from the surrounding cloud core, or protostellar envelope. Within the envelope the temperature and density increases towards the center due to stellar heating. To preserve angular momentum some of the accreting material spreads out into a disk, which simultaneously serves to funnel matter onto the star. Angular momentum is also removed from the system through the launch of protostellar outflows and jets (Fig. \ref{fig:shu}c).

As the protostellar system evolves, more and more mass is found in the star and disk compared to the remnant envelope. The envelope is finally dispersed on time scales of $\sim$1 Myr, leaving the a pre-main sequence star and a Keplerian disk (Fig. \ref{fig:shu}d). The change of name from protostar to pre-main sequence star signifies that the central star has become hot enough for fusion. The circumstellar disk is often referred to as a proto-planetary or planet-forming disk to signify that these disks are where planets assemble \citep{Williams11}. Recent observations suggest, however, that planet formation may begin much earlier, already at the protostellar stage \citep{ALMA15,Harsano18}.

The protoplanetary disk stage lasts for $\sim$1--10 Myrs based on disk occurrence rates in stellar clusters with known median ages \citep{Mamajek09}. During this time the disk material is accreted onto the star, onto planets, and dispersed through interactions with stellar photo-evaporative winds \citep{Ercolano14}. What is left behind is a nascent planetary system that can continue to evolve for 100s of Myrs due to collisions between remaining planets and planetesimals. No more gas can be added to planets at this point, and the formation time scale for Gas Giants is therefore set by this protoplanetary disk gas dispersion time scale.  For a discussion of astrophysical and solar system constraints on gas dispersion timescale the reader is referred to \citet{Pascucci10}.

\begin{figure}[htbp]
\begin{center}
\includegraphics[width=\textwidth]{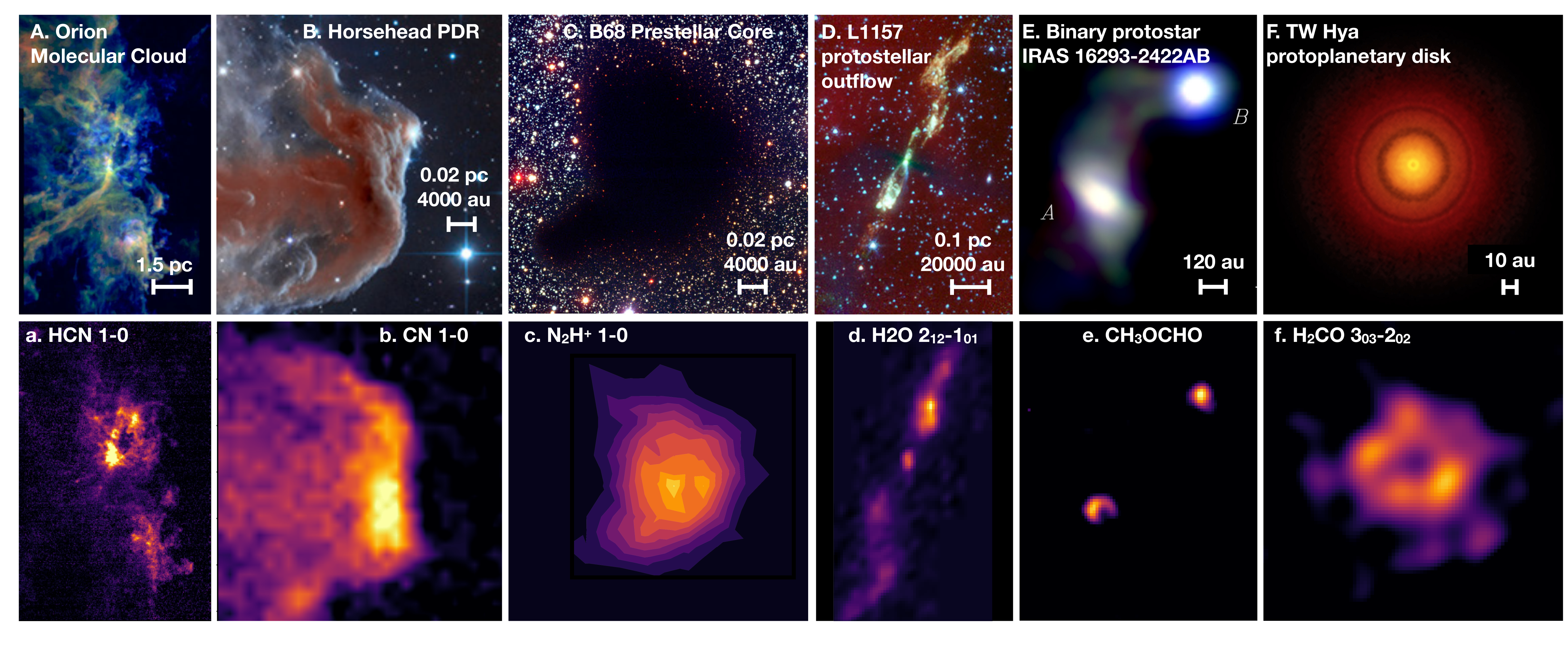}
\caption{Illustrations of characteristic chemical structures associated with the different stages and scales of star and planet formation. A. The Orion Molecular Cloud (OMC) as seen in CO emission (Image credit \& copyright J. Pety, the ORION-B Collaboration \& IRAM). a. The OMC in HCN 1-0 emission (data from \cite{Pety17}). B. The Horsehead photondominated region (PDR) (Image credit: Robert Gendler; ESO, VISTA, HLA, Hubble Heritage Team (STScI/AURA)). B. CN $1-0$ towards the Horsehead PDR (data from \cite{Guzman15}). C. The B68 prestellar core (Credit: ESO). c. B68 in N$_2$H$^+$ $1-0$ emission (data from \cite{Bergin01}). D. The outflow L1157 in infrared emission (Image credit: NASA/JPL/Spitzer). d. The same outflow in H$_2$O 179 $\mu$m line emission \citep{Nisini10}. E. The binary protostar IRAS 16293-2422 in dust emission at 1~mm, and in e.  in emission from a strong CH$_3$OCHO line from \citet{Jorgensen16}. F. Protoplanetary disk TW Hya from \citet{Andrews16}. The same disk in H$_2$CO line emission from \citet{Oberg17}.}
\label{fig:sf.chem}
\end{center}
\end{figure}

Each of the above evolutionary stages are characterized by a unique set of density, temperature and radiation structures, and a unique chemical starting point. It is therefore not surprising that while molecular lines are associated with all stages of star and planet formation, the kind of molecule that is observable changes during the star and planet formation process. Figure \ref{fig:sf.chem} shows the emission pattern of six characteristic molecules commonly used to explore cloud structure (HCN), PDRs (C$_2$H), prestellar cores (N$_2$H$^+$), protostellar outflows (H$_2$O), protostellar cores (complex organics), and protoplanetary disks (H$_2$CO). 

\subsection{Chemical reactions in interstellar and circumstellar environments}
\label{sec:Reactions}

Before considering the chemistry of specific regions associated with the formation of planetary systems it is useful to consider the chemistry most of them have in common. Even the densest phases of star and planet formation are rarified compared to planetary atmospheres. With a few exceptions they are too rarified to allow for three-body reactions of the kind $A+B+C \rightarrow ABC* \rightarrow AB+C$, where three species collide on timescales shorter than the dissociation timescales of an excited molecular complex, and the $A-B$ bond formation energy is carried away by $C$. The lack of three-body reactions in space limits what kind of bond formation can occur \citep{Herbst73}. A second important limitation is that most of the environments we consider in this review are cold, and atoms and molecules therefore lack sufficient kinetic energy to overcome substantial reaction barriers, though tunneling can mitigate this for some types of reactions \citep[e.g.][]{Hasegawa93,Cazaux11,Hama15}. 

\begin{figure}[htbp]
\begin{center}
\includegraphics[width=0.8\textwidth]{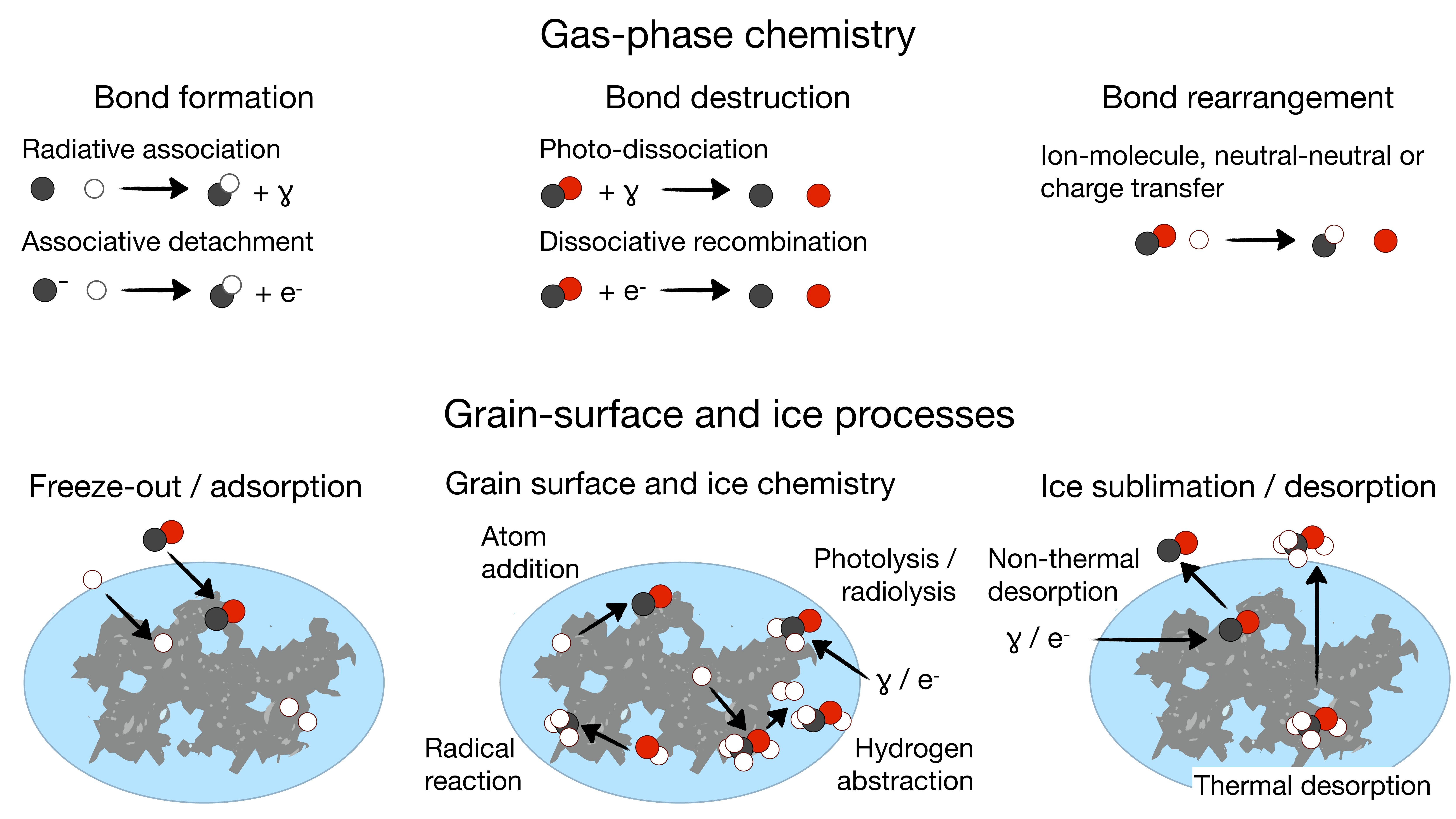}
\caption{Common chemical reaction types in interstellar and circumstellar media. The top row lists reactions in the gas-phase that leads to bond formation, destruction and rearrangement, respectively, while the bottom row focuses on grain surface reactions and grain-gas interactions.}
\label{fig:reactions}
\end{center}
\end{figure}

With these limitations in mind the top panels of Fig. \ref{fig:reactions} illustrate  the most common bond formation, destruction, and rearrangement reactions in astrochemical gas-phase chemistry networks.
In the absence of three-body reactions, bonds form  through radiative association and associative detachments, where the bond formation energy is carried away by photons and electrons, respectively. Under most interstellar conditions, radiative association is the most important gas-phase bond formation pathway. Even so, for many reactions of the form $A+B\rightarrow AB+ {\rm h\nu}$ a prohibitory number of collisions is required for a bond formation event to take place,  because collision and dissociation are fast ($\tau_{\rm diss}\sim10^{-13}$ s) compared to vibrational transitions ($\tau_{\rm rad}\sim10^{-3}$ s), corresponding to a very low effective reaction rate. The  rate is higher for reactions with accessible electronic transitions, which  reduces $\tau_{\rm rad}$, and with entrance energy barriers or large products, both which increases $\tau_{\rm diss}$. Still radiative attachment tend to only be important for reactions where the reactants are very abundant and thus collisions plentiful, which in practice implies that one of the reactant is the most common element in the Universe, i.e. hydrogen in either atomic or molecular form \citep{Herbst73}. Compared to radiative association, associative detachment reactions are fast, but it plays a small role due to that it requires one of the reactants to be an anion and anions are generally rare \citep{Millar07}.

Molecular bonds are destroyed through absorption of photons (photodissociation), collision with electrons (dissociative recombination), and in shocks through high-velocity collisions. Of these photodissociaton dominates in UV exposed interstellar and circumstellar regions. Photodissociation is the process through which a molecular bond is cleaved following the absorption of a photon: $AB+h\nu\rightarrow A+B$. Typical covalent bond strengths are above $>$5 eV and UV photons are thus required. Most molecules can be directly photodissociated through the absorbance of a photon into a dissociative electronic state, and therefore by all photons above some energy threshold. H$_2$ and some other small molecules are dissociated through absorption of photons at specific frequencies,  i.e. by resonance line photons. In these cases the photon absorption initially excites the molecule into an excited bound state, which can either cross over into a dissociative state (predissociation), or spontaneously decay into a dissociative state (spontaneous radiative dissociation) \citep{vanDishoeck88,Lee96}. For molecules that are only dissociated at discrete frequencies and abundant (especially H$_2$ and CO), self-shielding is often more important than shielding by dust. Photodissociation in an astrophysical context is described in detail by  \citet{Hollenbach99} and photodissociation rates and cross sections are found in \citet{Heays17}.

The final group of gas-phase reactions are those that result in a bond rearrangement, i.e. reactions of the type $AB+C\rightarrow A+BC$. Note that in this case there is no need to radiate away excess energy, since the net bond formation energy is carried away by $A$ as kinetic energy. These reactions typically involve radicals or ions or both. The importance of ions for molecule formation and evolution in the dense interstellar medium was noted in the early 1970's by \citet{Herbst73}. Ion-molecule reactions are typically exothermic, required at the low temperatures of most regions of interest to us, and reactions between ions and symmetric molecules with no permanent dipoles have a temperature-independent rate coefficient due to long-range charge interactions via an induced dipole which is modeled using the Langevin rate: $2\pi e \sqrt{\alpha/\mu}$ \citep[for a complete chemical reference:][]{levine09}.   Here $\alpha$ is the molecular polarizability and $\mu$ the reduced mass.  Higher, temperature-dependent, rates are found in the presence of a permanent dipole for the neutral molecule \citep{Adams85}. 

In diffuse media, and at the edges of dense clouds, surfaces of disks and other exposed regions, ions are produced through photoionization: $X^r+h\nu \rightarrow X^{r+1} + e$. The widespread presence of hydrogen atoms within the Galaxy means that UV photons with energies that exceed the hydrogen ionization threshold of 13.6 eV ($\lambda < 912$~\AA ) are largely absent. Phototionization of H and other atoms, and its balance with recombination is reviewed by \citet{Ferland03}. In dense interstellar and circumstellar environments, the central molecule powering the gas phase chemistry is H$_3^+$, which is a product of cosmic ray or X-ray ionization of H$_2$  \citep{Graedel82, Herbst89, Millar91, Cleeves13}. The H$_3^+$ density generally regulates the gas-phase chemical timescale, and because the space density of H$_3^+$ is close to constant with density, so is the overall gas-chemical timescale \citep{Lepp87}.

Though ion-molecule reactions are central in astrochemistry networks, there is also a large  number of neutral-neutral reactions that have significant import. For example, the formation of water in hot ($T > 400$~K) gas is powered via O $+$ H$_2$ $\rightarrow$ OH $+$ H followed by OH $+$ H$_2$ $\rightarrow$ H$_2$O $+$ H \citep{Wagner87}.
A more general discussion of astrochemical reactions can be found in \citet{Herbst95} and \citet{Wakelam10}. In the past couple of years there has been increasing realization that neutral-neutral reactions govern some of the cooler astrochemistry, including the formation of some organic species \citep{Shannon13,Balucani15}.

A completely different set of chemical considerations are related to the presence of interstellar dust grains \citep[e.g.][]{Hasegawa92,Hasegawa93,Garrod08}. Grains are rare compared to gas-phase molecules, but play a large role in regulating  astrochemical structures. First they act as sinks of gas-phase atoms and molecules. Grains are cool ($<$30~K) in most star and planet forming environments, and $<$10~K in prestellar cores \citep{Crapsi07,Pagani07}. Atoms and molecules that collide with a cold grain, with the exception of H$_2$ and He, will adsorb or freeze-out/condense\footnote{In the strict chemical sense, given the low pressures of interstellar space, the physical process involved for the vapor to solid transition is deposition.  However, in the astrophysical and cosmochemical literature it is common to label this transition as condensation.} onto the grain. Together with chemical reactions between adsorbed molecules and atoms on the grain surfaces, this freeze-out process results in the build-up of icy grain mantles.

The adsorption of gas-phase atoms and molecules can be described as a grain-atom/molecule collision as long as the sticking probability is high. The gas deposition timescale (and any timescale that depends on deposition/freeze-out) therefore has an inverse dependence on density:

\begin{equation}
    \tau_{\rm gas-grain} = 1/{ x_{\rm gr}n_{\rm H_2}\sigma_{\rm gr} v_{i,gas}}\;\;
\label{eqn:tau_dep}
\end{equation}

\noindent In this equation $x_{gr}$ is the abundance of cold grains, $\sigma_{gr}$ their surface area, and v$_{\rm i,gas}$ the thermal velocity of the particular molecule $i$.   Interstellar grains are characterized with a size distribution that follows an powerlaw of size $a$ distributed by $a^{-3.5}$ with sizes ranging from large molecules ($\sim$10 \AA) to large grains with $a \sim$0.25 $\mu$m \citep{MRN77}. Thus there is more surface area in small grains, while the mass resides in the larger grains \citep{Draine95}.  

Once atoms/molecules are on the ices, the initial chemistry is dominated by H atom addition  \citep{Hasegawa92}. It has long been known that the most abundant molecule in space, H$_2$, is required to form via grain surface catalytic chemistry \citep{Gould63, Hollenbach71,Wakelam17}. H atom additions to heavier atoms such as O leads to the formation of water and other saturated species up to methanol (CH$_3$OH) in size \citep{Tielens82}. Note that these reactions are effectively three-body reactions, with the grain acting as the third body, adsorbing the bond formation energy; i.e. the grain acts as a catalyst. Once CO, the most abundant molecule produced through gas-phase chemistry, begins to freeze out, there is also an important CO-mediated chemistry to form e.g. CO$_2$.   

\begin{figure}[htbp]
\begin{center}
\includegraphics[width=0.8\textwidth]{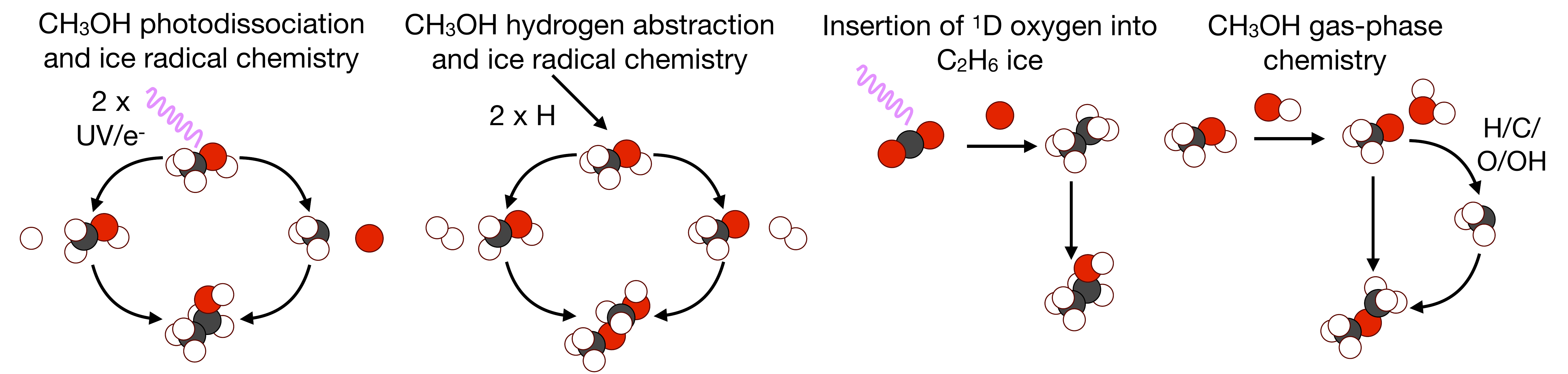}
\caption{Four proposed pathways to forming O-rich organics during star and planet-formation. The multitude of pathways and a similar diversity in pathways to hydrocarbons and N-containing complex organics probably explains why complex organics are present in many kinds of interstellar and circumstellar environments. Note that a majority of the displayed chemistry is barrierless, but e.g. the radical ice chemistry is still faster at elevated temperatures because of faster radical diffusion through the ice. }
\label{fig:com-chemistry}
\end{center}
\end{figure}

Reactions on grain surfaces and in icy grain mantles can also produce more complex molecules. Figure \ref{fig:com-chemistry} shows four proposed and experimentally verified pathways to complex organics originating in the ice mantles of interstellar grains. Radicals produced via UV or electron-mediated dissociation, or via hydrogen addition and abstraction, can recombine to produce more complex molecules \citep{Garrod08,Bennett07a,Oberg09d,Chuang17}. These reactions can proceed at all temperatures between neighboring radicals, and throughout the ice if the grain is heated to above $\sim$25~K, enabling radical diffusion. Oxygen insertion into hydrocarbons produces alcohols and aldehydes \citep{Bergner19a} in ices as cold as 10~K, and may produce complex organics in low-temperature environments. Finally, sublimated methanol ice can seed a complex gas-phase chemistry at a range of gas temperatures \citep{Charnley92,Balucani15,Vasyunin17}. 

Grain surface and ice species sublimate through thermal desorption, photodesorption and electron stimulated desorption, release of chemical energy during a chemical reaction, grain-grain collisions and other energetic events. Thermal desorption tends to dominate the grain-gas balance in protostellar envelopes and other heated regions \citep{Leger85}, while photodesorption regulates the gas-grain balance in UV exposed regions \citep{Hollenbach09}. In cold, UV-shielded regions, desorption through release of chemical energy is suspected to be the dominant desorption mechanism \citep{Garrod07}. Thermal desorption has been characterized for most common grain surface species \citep[e.g.][]{Fraser01,Collings04,Oberg05}, and found to be quite complex; the desorption barrier of some species is very sensitive to the details of the ice matrix \citep[e.g.][]{Fayolle16}.  Based on information from laboratory, and the general physics behind physical adsorption, we can classify volatiles based upon their sublimation temperatures.   CH$_4$, CO and N$_2$ have the highest volatility with sublimation temperatures near 20-40~K (depending on the surface and pressure); these can be classified as hypervolatiles.  Other species, such as water, CH$_3$OH, CO$_2$, and NH$_3$ can be denoted as volatiles.  Finally, of course, silicates and carbonaceous grains have the lowest volatility and are classified as refractories.

Of the different non-thermal desorption mechanisms, photodesorption has been most extensively explored via laboratory experiments in the past decade.
 Early experiments demonstrated that UV photodesorption is highly efficient \citep{Westley95,Oberg07b, Oberg09a,Oberg09b}, while molecular dynamics simulations \citep{Andersson06,Andersson08} revealed that the process can be quite complex and occur through several different channels, even for the same molecule. Most of the theoretically predicted channels have since been seen in experiments \citep{Fayolle11,Bertin13}. More recently, experiments on chemical desorption have begun to constrain this important process \citep[e.g.][]{Minissale16}.

\subsection{Methods of astrochemistry}

\subsubsection{Astronomical observations}

The nearest active star and planet forming regions are $\sim$100 pc away and remote sensing is therefore the only option to obtain direct, empirical evidence of the chemistry that impacts planet formation. An important exception to this principle is the solar system, where the chemical composition of comets, asteroids, planets, and moons, all encode information about their formation conditions. Missions to comets, asteroids, and planets, along with remote observations of these same bodies, have provided invaluable information about their chemical make-up, which has then been used to map out the origins of their different constituents. For example the elevated D/H ratio found in terrestrial, asteroid and cometary water has been used to demonstrate that their water reservoirs to a large extent are inherited from the cold interstellar medium \citep{Cleeves14}.

Yet most astrochemical observations are remote. A central limitation of these remote observations is the fact that H$_2$ is unemissive at the  temperatures and densities characteristic of mass reservoirs throughout the phases of star and planet formation \citep{Evans03}.  To determine chemical abundances in each phase requires the use of calibrated probes, such as CO isotopologues and optically thin thermal continuum emission from dust grains, to determine the hydrogen content and therefore the total gas mass \citep[for greater discussion see,][]{Bergin17}.  More broadly, beyond H$_2$ and CO, there are a host of detectable emission lines arising from electronic, vibrational, torsional, and rotational states of hundreds of species, which together can be used to probe the physical and chemical characteristics of specific regions.

In principle, astrochemical observations are possible at all wavelengths where  molecules emit and absorb photons at discreet energies, i.e. from UV to radio wavelengths. In practice the vast majority of astrochemical observations are, however, carried out at longer wavelengths, at infrared (IR), far-IR or Terahertz, sub-millimeter and millimeter, and radio wavelengths. Molecular spectral features at near and mid-IR wavelengths are generated by molecular vibrations, and are used to probe both gas-phase and solid-state compositions \citep{vanDishoeck98,vanDishoeck06b, Pontoppidan14, Boogert15}. For gas-phase molecules, most studies have focused on warm enough gas that the lines are in emission. Solid-state compositional studies, are by contrast mainly carried out in absorption because ice mantles would sublimate before they can be heated enough to emit at near to mid-IR ($<$~20$\mu$m) wavelengths \citep[e.g.,][]{Boogert15}; emission is detectable at longer wavelengths \citep{McClure15, Min16, Kamp18}. Some IR observations can be carried out from the ground, but complete IR spectral coverage is only possible from space, and the most recent IR mission is the Spitzer Space Telescope. Far-IR or Terahertz spectroscopy are exclusively possible from space (e.g. the Herschel Space Telescopes), or very high altitudes (SOFIA). These wavelengths give access to the fundamental rotational transitions of light hydrides such as H$_2$O \citep{vanDishoeck14}, and HD \citep{Bergin13}.

Moving to longer wavelengths, submillimeter and millimeter observations are possible from the ground, but the shorter wavelengths require exceptionally dry locations, motivating the placement of the Atacama Large Millimeter and submillimeter Array (ALMA), in the Atacama dessert. Most small and mid-sized molecules, including CO \citep{Wilson70}, HCN and CH$_3$OH, have their fundamental transitions in this wavelength regime, and a majority of molecular observations make use of this wavelength regime. These transitions are sufficiently low energy to be excited even in the coldest interstellar and protoplanetary disk conditions, where temperatures are $<$20~K. The longest wavelengths, beyond 3~mm enable accurate observations of large organic molecules \citep[e.g.][]{McGuire18a}. Through these different kinds of astronomical observations more than 200 molecules have been identified in interstellar and circumstellar environments, the vast majority of which are organic \citep{McGuire18b}.

Interpreting these observations relies on detailed radiative transfer codes, and other methods to estimate molecular abundances and excitation temperatures from molecular line observations. The reader is referred to the following references: \citet{Hogerheijde00}, \citet{Brinch10}, \citet{vanderTak11},\citet{Dullemond12}, and \citet{Shirley15}

\subsubsection{Astrochemical models}

Our theoretical understanding of astrochemical processes rely on a range of modeling techniques, each of which makes a different compromise between computational accuracy and computational speed. The different theoretical tools are reviewed in detail by \citet{Garrod13b}, and \cite{Cuppen13}, and are briefly summarized here. The most computational expensive models are {\it ab initio} quantum mechanical calculations, which are employed in astrochemistry to calculate photodissociation cross sections, collisional excitation coefficients, reaction potentials, and some molecule-grain surface interactions \citep[e.g.][]{Heller78,Green78,vanDishoeck88, Schoier05}. Molecular dynamics (MD) simulations use these reaction potentials to calculate how molecules react on pico-second timescales using classically calculated trajectories \citep[e.g.][]{Andersson06}.  Within astrochemistry, MD simulations has been used to uncover ice reaction mechanisms that can then be parametrically incorporated into higher-level modeling efforts.

Longer simulations require that reaction probabilities are parameterized, which is done through so called microscopic or kinetic Monte Carlo models. In these precise kinetic models, the motions of individual atoms, radicals and molecules are followed on top of and inside of ice mantles based on calculated probabilities. These models are typically employed to investigate specific grain-surface chemical processes, such as H$_2$ or CH$_3$OH formation. For grain surface processes this detailed treatment is important, since only considering the chemistry statistically can yield missleading results \citep[e.g.][]{Tielens82,Cuppen09}. For larger networks consisting of more than a handful of species and 10s of reactions, microscopic Monte Carlo simulations become intractable. 

The next level of theory consists of macroscopic stochastic or rate equation models, both which are routinely employed to carry out comprehensive models of gas-phase and solid-state interstellar chemistry. 
In macroscopic Monte Carlo models,  the master equation for the combined gas-phase and surface
chemistry is solved using Monte Carlo techniques \citep{Tielens82, Charnley98,Vasyunin09}. These models employ averaged grain-surface reaction rates, and couple them with gas-phase chemistry, producing exact gas-phase and grain-surface populations for each chemical species, under the assumption that the detailed ice structure captured in the microscopic models is not important.

The most comprehensive astrochemistry models, in terms of number of species and reactions, exclusively use the rate equation approach or some modification of it. In these models a set of ordinary differential equations that describes the gas-and grain-phase chemical kinetics is solved to produce time dependent abundance information for each chemical species as the gas and ice are exposed to constant or changing physical conditions.  Over the past 20 years detailed chemical networks have been constructed that are now  widely applied to  chemistry of dense cloud cores, planet-forming disks, and circumstellar envelopes \citep{UMIST12, KIDA14}. Rate equation models have evolved in sophistication to compete in accuracy with macroscopic Monte Carlo methods \citep{Garrod08b}. One important feature of most contemporary rate equation models is the treatment of ice surfaces separate from ice layers \citep{Hasegawa93}, and so called three-phase models (ice mantle, ice surface and gas-phase) are now standard \citep{Vasyunin13,Garrod13}. In addition, \citet{Cuppen17} provide a detailed discussion of the inclusion of grain surface chemistry into networks.

The final class of chemical models are ones that link the gas-phase/gas-grain chemical evolution to the thermal properties of the gas and dust.  Atomic \citep{Dalgarno72} and molecular \citep{Goldsmith78} emission are primary coolants for the gas in regions where the density is below where the gas and dust temperatures are decoupled; this occurs at densities of $\sim 10^5$~cm$^{-3}$ for interstellar dust mixture and size distribution \citep{Burke83}. Thus the chemistry can be intimately coupled to the resulting gas temperature.  This class of models are labelled as thermochemical models and have been applied to the general interstellar medium \citep{Wolfire95}, dense clouds exposed to enhanced radition fields \citep{Tielens85, Sternberg95, Kaufman95, LePetit06}, and protoplanetary disks \citep{Gorti04,Woitke09, Kamp10, Bruderer12, Du14}.

\subsubsection{Laboratory experiments}

Most astrochemical processes cannot be calculated from first principles with sufficient precision. Rather, astrochemical models rely on laboratory experiments to anchor calculations of spectra of interstellar and circumstellar molecules, and of molecular and atomic excitations, and to predict the outcome of astrochemical reactions.

{\it Spectroscopy.} Molecular spectra cannot be calculated {\it a priori} to the precision required for identifications of interstellar spectral lines with specific carriers. Laboratory measurements of molecular spectra across the electromagnetic spectrum is a cornerstone for the astrochemical project \citep{Widicus-Weaver19}. Spectral measurements of stable molecules are relatively straightforward in the gas-phase. Yet data-bases are far from complete, especially for isotopologues of common interstellar molecules \citep{muller05}.  Spectra of radicals and ions are more cumbersome to obtain since they require the ionization or dissociation of a precursor. This produces a mixture of species due to multiple dissociation products and/or subsequent chemistry, and the resulting spectra need to be carefully analyzed to identify the lines belonging to the target molecule \citep[e.g.][]{McCarthy06,Bizzocchi17}. 

Spectra of ices are acquired in high or ultra-high vacuum chambers where thin (nm--$\mu$m) ices are vapor-deposited before measuring their infrared or far-infrared spectra \citep[e.g.][]{Hagen83}. Ice spectroscopy differs from gas-phase spectroscopy in one important aspect, which is that the spectral band position, shape and strength depends on the nature of surrounding molecules. Detailed spectroscopic studies of the target molecule in different ice environments is therefore needed to characterize ice mantles in space \citep[e.g.][]{Gerakines95, Ehrenfreund96}. There are multiple gas-phase and solid-state spectral databases serving the astrochemistry community, including the JPL spectral database, the Cologne Database for Molecular Spectroscopy (CDMS), Splatalogue, the Leiden ice database (http://icedb.strw.leidenuniv.nl) and the NASA Goddard Cosmic Ice Laboratory IR spectra database \\(https://science.gsfc.nasa.gov/691/cosmicice/spectra.html).\\

{\it Molecular Excitation.} 
Molecules can relax and release quantized amounts of energy via rotation/vibrational motions and electronic excitation.    Transitions associated with rotational motions are powered by a rotating dipole corresponding to the lowest change in energy of tens to hundreds of Kelvins and are found at ($\Delta E$ = hc/$\lambda$)  mm/sub-mm wavelengths, while vibrational modes have $\Delta E \sim$0.1 eV or 1000 K, observed in the infrared, and electronic transitions occur $\sim 10$eV or 100,000 K corresponding to UV wavelengths. In star and planet forming regions most of the mass is at cold (T $<$ 100~K) temperatures, where only rotational transitions are excited. IR lines are also important, however, and are observed in absorption in cold regions, and in emission close to young stars. The reader is referred to the textbooks of the field to explore these questions more thoroughly from perspective of molecular physics \citep{herzberg_v1, Townes55, Gordy70}.

In interstellar and circumstellar regions of interest to this review, the emission of atoms and molecules is generally excited via collisions with molecular hydrogen with temperature-dependent excitation cross-sections ($<\sigma v>$) that are specific to H$_2$ \citep[for details see, ][]{Flower12}.  In the densest interstellar and circumstellar regions, the details of the excitation process are not important, since local thermal equilibrium (LTE) can be assumed. In many, if not most, 'dense' astronomical regions the densities are not sufficient for LTE to hold, however and the collision rates are key to use molecular emission to derive molecular abundances, as well as to use molecular emission as tracers of gas densities and temperatures. The collision rate for molecule X is n$_X$n$_{H_2}<\sigma v>$, and thus depend on H$_2$ space density, gas temperature and the collisional excitation cross section.  For detailed discussion of this topic, and the methods used to determine the H$_2$ density and temperature, along with molecular abundance, the reader is referred to \citet{Evans99}, \citet{Goldsmith99}, \citet{Shirley15}, and \citet{Mangum15}. The different techniques that have been developed to obtain collisional excitation data are summarized in a review by \cite{Smith11}. \\

{\it Gas-phase reactions.} \citet{Smith11} also reviews experimental techniques for characterizing astrochemically relevant gas-phase reaction rates, and  product branching ratios. The experimental apparatus used to obtain these depend on the reaction under investigation. Ion-neutral reactions where the neutral is a stable molecule are relatively straightforward to quantify using ion traps and flow tubes, and the experiments tend to agree with theoretical calculations within a factor of 3. Two important developments of the past few decades are the so called Selected Ion Flow Tube method where the reactant ion is mass selected and the reaction rate can therefore be measured with great precision \citep{Martinez08}, and the CRESU apparatus, which enables the measurements of reaction rates at very low temperatures due to rapid, supersonic expansion of reactant gas \citep{Sims93}. Low temperature measurements are important since it is difficult to predict the temperature dependence of reaction rates at low temperatures based on room temperature experiments.

Neutral-neutral reactions are more challenging to characterize, especially between unstable species, due to difficulties in mass-selecting neutral molecules and radicals. A small number of  reaction systems have, however, been explored using either the flow tubes and the CRESU method. A key discovery from these experiments is that the rate coefficients of many neutral-neutral reactions increase with decreasing temperature, often in contradiction with pre-existing calculations. This characteristic has resulted in a re-evaluation of the importance of neutral-neutral reactions in cold interstellar and circumstellar environments \citep{Smith11}.\\

\begin{figure}[htbp]
\begin{center}
\includegraphics[width=0.8\textwidth]{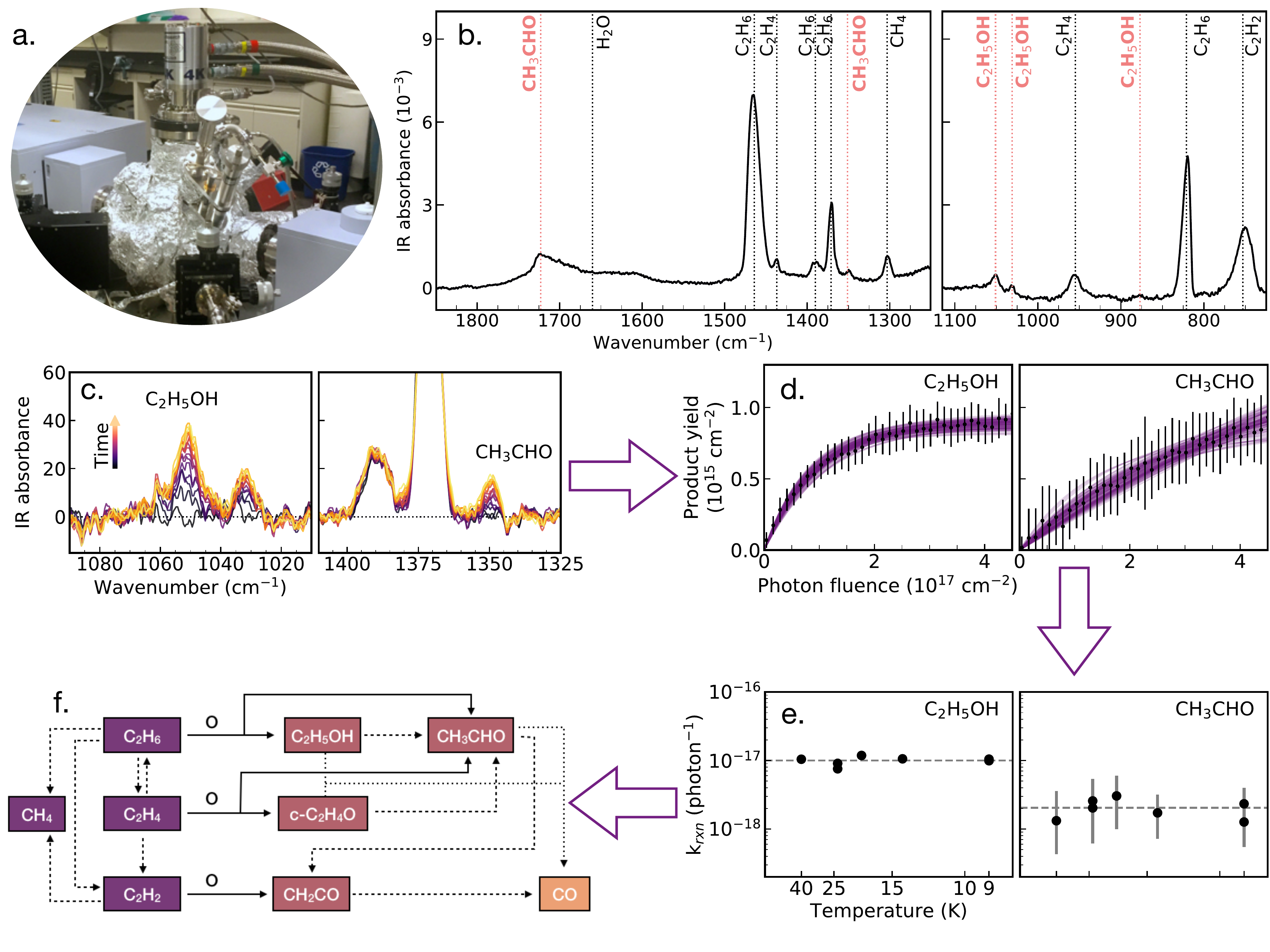}
\caption{Flowchart for a typical astrochemistry solid-state experiment, using data from \citet{Bergner19a}. Experiments are carried out in vacuum chambers (a), where an initial ice is deposited on a substrate. The ice is exposed to atoms, heat or energetic processing and the chemical products are identified using spectroscopy (b), or mass spectrometry following ice desorption. If the ice products are monitored over time (c), product growth curves can be modeled (d) to obtain rate coefficients. By comparing rate coefficients obtained at different ice temperatures, the reaction barrier can be determined (e). This quantitative information, together with a qualitative identification of reaction products in different ice experiments can then be used to assemble astrochemically relevant reaction networks (f).}
\label{fig:lab-coms}
\end{center}
\end{figure}

{\it Solid-state reactions.} The chemical abundances on the surface of interstellar grains depend on a combination of adsorption, desorption, diffusion and surface chemical reactions. These are characterized using surface-science apparatuses designed to mimic the high vacuum and range of temperatures (down to 4~K) characteristic of interstellar environments. Three kinds of experimental domains can be identified, those dealing with warm grains where chemisorption is important, those dealing with bare grains, and those dealing with ice covered grains. The first kind of experiment addresses e.g. H$_2$ formation in the warm, diffuse interstellar medium, where the residence time of any atom would be too short unless chemical binding between the incoming atom and the surface is invoked. 

In the first family of experiments,  graphite analogs are exposed to (H) atoms followed my mass-spectrometric or microscopic monitoring of reaction kinetics \citep[e.g.][]{Hornekaer08}. Experiments on cold bare grains are most commonly simulated by exposing cryo-coooled silicate and graphite surfaces to atomic beams followed by temperature programmed desorption experiments (TPD) where the surface is rapidly heated up while the chamber is monitored mass-spectrometrically \citep[e.g.][]{Katz99}.  Finally, as illustrated in Fig. \ref{fig:lab-coms}, experiments aimed at understanding the chemistry of ice covered grains typically proceed by exposing a cryo-cooled metal or glass surface to a molecular beam of water vapor or another common interstellar gas (mixture) to build up a thin (nm to 10s of nm) ice. This ice is then exposed to atoms, UV photons, electrons, X-rays, energetic particles or heat, and ice desorption, restructuring, and chemistry is monitored using infrared spectroscopic and mass-spectrometric techniques \citep[e.g.][]{Watanabe02,Oberg15}. 

\section{Setting the Chemical Trajectory: The Chemistry of Planet Formation Begins in Molecular Clouds and Cloud Cores}
\label{sec:early}

There is growing evidence  that the chemistry present in planet-forming disks is intimately linked to the coupled physical and chemical processes that regulate chemistry in the interstellar medium. It is in interstellar clouds, and even during their assembly, that the major volatile carriers of oxygen, carbon and nitrogen form: H$_2$O, CO, and N$_2$. These carriers remain the major ones through the later star and planet-forming phases, and their relative volatility and chemical reactivity are key to understand how the chemistry evolves during planet formation. The first objective of this section is to review why so much oxygen, carbon and nitrogen become incorporated into just these three carriers. The volatile reservoirs of two other elements, sulfur (S) and phosphorous (P) , should also be established in the dense cloud stage. Our understanding of the S and P chemistry is much less mature, however, and apart from a short section in \S~4.4, we do not treat these two prebiotically important elements in detail in this review. The reader is instead referred \citet{Vidal17,Vastel18,Laas19,LeGal19} and \citet{Lefloch16,Ziurys18,Jimenez-Serra18,Bergner19,Rivilla20} for some recent developments in our understanding of S and P astrochemistry, respectively.

Molecular clouds and cloud cores are also the formation sites of the most abundant volatile organics during planet formation -- CH$_4$ and CH$_3$OH \citep{Mumma11} -- as well as the pre-biotically interesting NH$_3$. These are the building blocks of any proceeding complex organic chemistry and their initial abundance may effectively limit how many large organic molecules can form in the later protostellar and protoplanetary-disk stages. The second objective of this section is to review the chemistry responsible for NH$_3$, CH$_4$ and CH$_3$OH production in clouds and how it relates to the formation of the major oxygen, carbon, and nitrogen carriers H$_2$O, CO, and N$_2$.

Finally, molecular clouds and cloud cores are sites of isotopic fractionation, the fingerprint of which is seen today in solar system water (as one example). Deuterium enrichments that can be linked to low-temperature fractionation processes are found in the Earth's ocean, as well as in cometary ices, and they represent our most important evidence connecting solar system chemistry to interstellar chemistry. There are several chemical processes through which molecules can become enriched or depleted in heavy, stable isotopes, and the final objective of this section is to review the physics and chemistry behind interstellar isotopic fractionation, especially how it pertains to deuterium fractionation in cold cloud cores. 

Before treating these specifics it is useful to consider the overall chemical structure of molecular clouds. In the molecular cloud stage, key quantities that influence the overall chemical trajectory are the radiation field (cosmic rays, X-rays, UV radiation), gas and dust temperature, and gas density. These quantities are both time and location dependent. As clouds compact they become denser, and therefore colder and less permeable to radiation, while in an already compact, dense cloud, the exposed edges are warmer and more irradiated than the shielded interiors. In both the time-dependent and cloud-depth dependent description of cloud chemistry, the chemical structures initially (or closer to the cloud surface) depend on an interplay of photon-mediated processes (photoionization, photodissociation, photon-heating, and photodesorption), and molecular formation through gas and dust surface chemical processes. In the more shielded cloud regions (or later times) the chemistry instead becomes regulated by a competition between gas-phase chemistry, and increasing freeze-out of molecules from the gas-phase onto dust grains.

\begin{figure}[htbp!]
\begin{center}
\includegraphics[width=0.8\textwidth,angle=0]{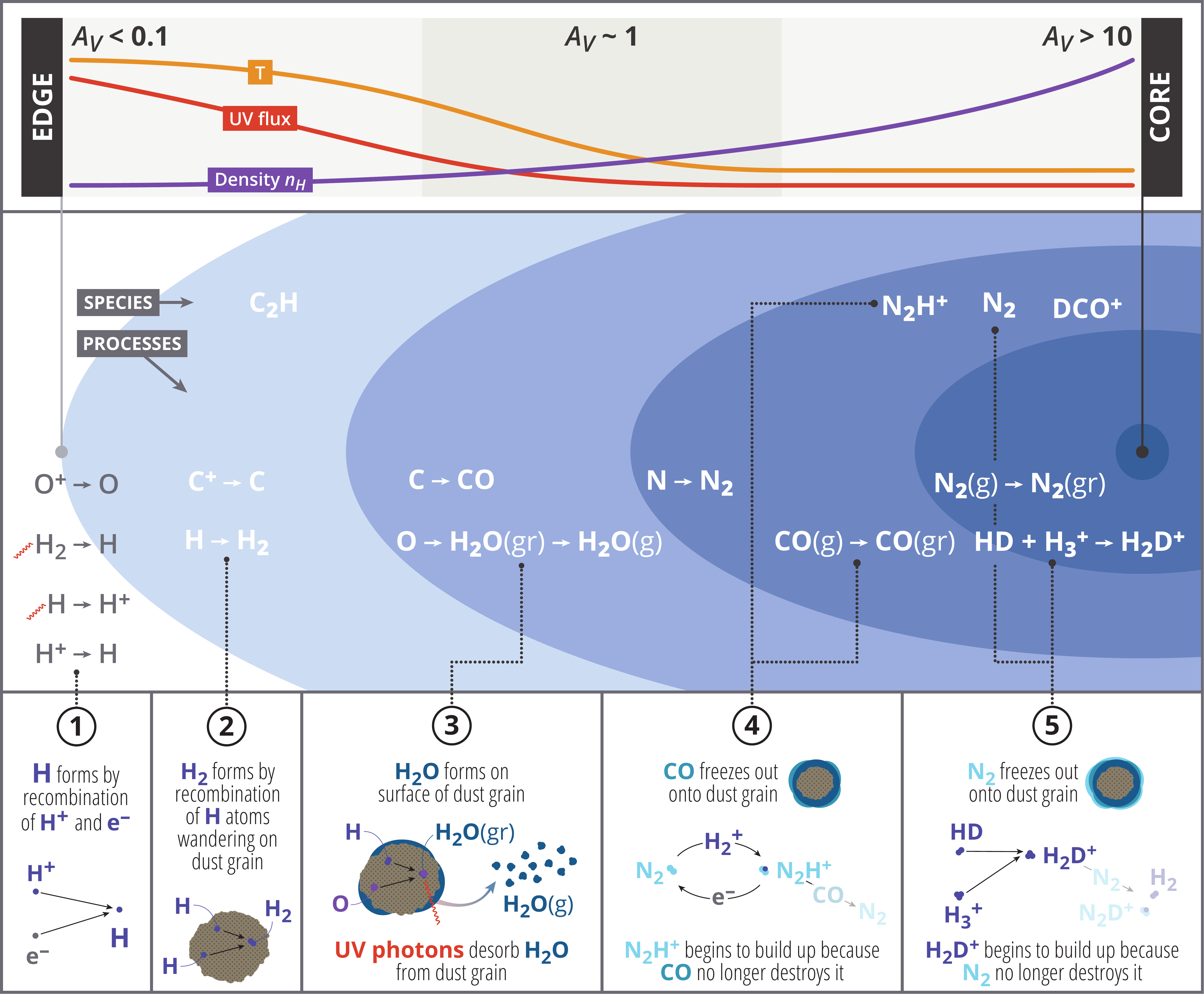}
\caption{Illustration of the chemical structure of a molecular cloud from the UV exposed cloud edge (left) to the protected cloud core (right), where cloud depth is parameterized by extinction (measured in visual magnitudes A$_{\rm V}$). At the cloud edge, the most energetic UV photons, capable of ionizing H and O have already become fully absorbed. In the outer cloud layers carbon transitions from atomic ions to neutral atoms, and hydrogen from atoms to molecules. In the inner surface layer carbon is rapidly converted to CO, effectively locking in most of the volatile carbon budget. In the same cloud layer the remaining oxygen is mainly incorporated into the much less volatile water. Most nitrogen is expected to become incorporated into the hypervolatile N$_2$, but a substantial amount may also be present as NH$_3$, already in the water formation zone. In the deepest molecular cloud region, all molecules except for H$_2$, H$_3^+$ and their isotopolgoes rapidly freeze-out onto grains, resulting in a volatile-depleted gas, and efficient deuterium fractionation. While this cartoon primarily  presents a static view of the cloud chemical structure, it also illustrates what happens as a function of time (going from left to right), when a package of diffuse cloud material compacts to form a dense cloud. Image credit: K. Peek, after the PDR illustration presented in \citet{Tielens85}.}
\label{fig:pdr-cartoon}
\end{center}
\end{figure}

Figure \ref{fig:pdr-cartoon} shows an illustration of the resulting chemical structures of clouds, where the x-axis reflects depth into a cloud. 
In astronomy, such UV-exposed molecular cloud edges are traditionally characterized as Photo-Dissociation Regions or PDRs \citep{Tielens85, Hollenbach99}.
When viewed edge-on, these regions display characteristic chemical layers associated with increasing UV attenuation. Indeed, Fig.~\ref{fig:pdr-cartoon} illustrates the chemical depth-dependent transitions for the most abundant elements (except for He), i.e. H/H$_2$, C$^+$/C/CO/CO(gr), O/CO+H$_2$O(gr), N/N$_2$/N$_2$(gr), where (gr) marks a species that has frozen out on a grain surface. The same kind of chemical transitions are expected in all interstellar and circumstellar regions that are externally illuminated, and classical PDR chemistry are therefore also considered model systems for understanding chemical layering in planet forming disks \citep{vanZadelhoff01,LeGal19b}. We note that  Fig.~\ref{fig:pdr-cartoon} would look qualitatively the same if the x-axis instead reflected time during cloud assembly and compaction \citep{Bergin04b, Glover07}. We can therefore use this classical PDR framework to explain the chemical trajectory present during cloud formation and core condensation, and in particular to address our three objectives to explain oxygen, carbon and nitrogen carriers throughout star formation, origins of common organics, and the presence of high levels of isotopic fractionation.

\subsection{Origins of Major Oxygen, Carbon and Nitrogen Volatiles: Photodissociation vs Formation}

The major volatile carriers of oxygen, carbon, and nitrogen form in molecular clouds. Why some species and not others end up dominating the oxygen, carbon, and nitrogen budget primarily depends on an interplay between photodissociation and gas-phase formation, and secondarily, how readily the products of this initial gas-phase chemistry can be converted into other molecules. 

\begin{figure}[htbp]
\begin{center}
\includegraphics[width=0.8\textwidth]{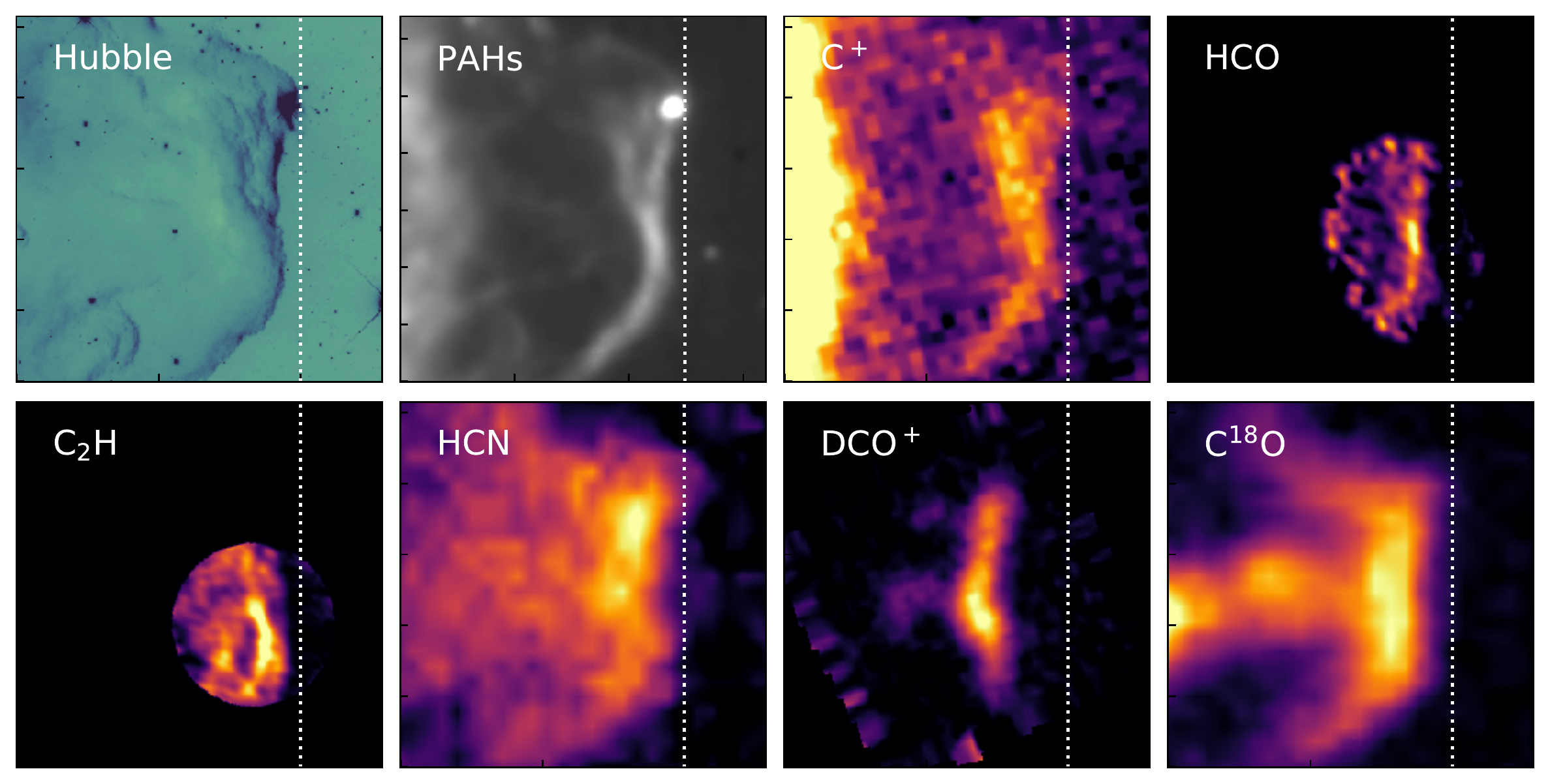}
\caption{Observations of the Horsehead nebula in infrared light from  HST's WFC3 infrared camera (Hubble legacy archive), PAHs \citep{Abergel03}, C$^+$ \citep{Bally18}, and five different molecular tracers: HCO, C$_2$H, HCN, DCO$^+$, and C$^{18}$O \citep{Gerin09,Pety05,Pety07,Pety17}. The dotted line shows the edge of the C$_2$H emission in all panels, denoting the edge of the molecular layer in the PDR. Note the different chemical layers probed by different species, i.e. C$^+$ and polycyclic hydrocarbons (PAHs) emit close to the surface, most molecules somewhat further into the cloud, and DCO$^+$ deep into the cloud core. Image credit: V. Guzm{\'a}n}
\label{fig:pdr-obs}
\end{center}
\end{figure}

As introduced above, the interstellar regions where photodissociation regulates the chemistry are referred to as PDRs. PDRs are most readily identified at the edges of dense clouds that face massive stars and are thus exposed to high levels of UV radiation. Indeed these are typically referred to as `Classical PDRs', and a famous example is the Horsehead Nebula, whose chemical structure is shown in Fig. \ref{fig:pdr-obs}. Molecular clouds that are not directly exposed to massive stars still always experience some level of irradiation and their outer structures too are determined by photoprocesses. In both scenarios, photons with energies higher than 13.6~eV become fully absorbed by atomic hydrogen at some distance from the cloud edge, resulting in a H$^+$/H transition (Fig. \ref{fig:pdr-cartoon}). Other species that require photons $>$13.6~eV, e.g. atomic oxygen, also transition to their neutral form at this boundary between a predominantly ionized and neutral medium. Species that have lower ionization energies, i.e. carbon, remain largely ionized. 

The second major transition is between H and H$_2$, which occurs when H$_2$ formation on grains out-competes H$_2$ photodissociation, which has a threshold of 11.2 eV. As described above, H$_2$ efficiently self-shields and this boundary is obtained when a certain hydrogen column is between the cloud and the ionizing source. This boundary marks the edge of the dense molecular cloud \citep{Narayanan08}. Concomitant with the selective shielding of UV photons by H$_2$, a combination of self-shielding and overall UV attenuation by dust results in a slow transition between C$^+$ and C. C is a starting point of an efficient oxygen-poor organic chemistry resulting in e.g. the build-up of C$_2$H in these outer cloud layers. At this point most oxygen, carbon, and nitrogen are in their atomic form, and what happens next is what determines the different chemical trajectories and reservoirs of these three elements.

Interior to the H-to-H$_2$ transition layer, CO formation becomes efficient enough to build up sufficient CO column density to self-shield against the still pervasive interstellar radiation field \citep{vanDishoeck88}. The ability of CO to self-shield effectively locks up most of the available carbon into CO, making it the main volatile carbon reservoir from this stage and onwards. The rapid conversion of carbon into CO, and the very slow conversion of CO into other molecules up until the disk stage, makes it an excellent tracer of gas mass, and it is indeed the most common tracer of gas mass through most stages of star and planet formation \citep[see][and references therein]{Bergin17}.

In the same PDR layer, oxygen is efficiently converted into water ice through hydrogen addition to oxygen atoms, and oxygen molecules on the surfaces of dust grains. This process is efficient enough that a majority of the available volatile oxygen goes into water ice, resulting in comparable water ice and CO vapor abundances \citep{Hollenbach09}. This process also starves the gas-phase of available oxygen which stunts the subsequent gas-phase formation of water vapor and O$_2$ \citep{Bergin00} as evidenced by measured low abundances of water vapor and systemic non-detections of O$_2$ towards nearly all\footnote{O$_2$ has surprisingly high abundance in one object, the $\rho$ Oph cloud \citep{Liseau12}.} lines of sight \citep{Snell00, Goldsmith00, Pagani03, vanDishoeck11, Caselli12, Wirstrom16}.

It is an interesting question why CO and H$_2$O become the dominant carriers of carbon and oxygen rather than say CH$_4$, which can form from hydrogen addition to atomic carbon on grain surfaces, and O$_2$, which can form through gas-phase chemistry. Some CH$_4$ certainly does form through grain-surface chemistry, but its abundance is low, a few \%, compared to water. The inefficiency of this chemistry can be understood from a delicate balance between rapid conversion of C into CO, just as radiation attenuation becomes high enough to allow for build-up of molecules on grain surfaces -- slightly closer to the cloud surface, where C is more abundant than CO, any molecule formed on the grain surface is rapidly photodesorbed and/or photodissociated. Some O$_2$ also likely forms in this PDR layer, but compared to CO it faces two important disadvantages: it does not self-shield efficiently and is therefore photodissociated deeper into the cloud, and O$_2$ is relatively easy to convert into water once it adsorbs onto a grain surface, resulting in water ice \citep{Ioppolo08}.

The case of nitrogen likely mirrors that of carbon, but with some important distinctions. Like CO, molecular nitrogen, N$_2$, forms in the gas-phase and it is expected, though not directly proven, that this is the major reservoir of nitrogen \citep{vanDishoeck93, Maret06, Daranlot12, Furuya18}. N$_2$ self-shields \citep{Heays14}, which limits its photodissociation, but because nitrogen has an intrinsically lower abundance than carbon, the self-shielding only becomes effective deeper into the cloud.   
In addition, N$_2$ formation chemistry is  initiated by slow neutral-neutral reactions \citep{PineaudesForets90, Gerin92}, and its formation is therefore slower than ion-molecule fostered production of CO. As a result, nitrogen  transitions from N to N$_2$ at deeper cloud layers. Once N$_2$ has formed it is chemically very stable, and effectively locks up the nitrogen through the remainder of the star and planet formation process. Because the nitrogen remains atomic for longer than carbon, a substantial portion of N may react with hydrogen on grain surfaces to form NH$_3$ \citep{Fedosev15}, analogous to water formation from atomic oxygen.  NH$_3$ is detected in the icy mantles of interstellar grains, but not at very high abundances \citep{Boogert15}. It has been suggested, however, that a substantial amount of the NH$_3$ is hidden as ammonium salts, and that together NH$_3$ and NH$_4^+$ may constitute an appreciable nitrogen reservoir \citep{Boogert15,Altwegg20}.

\subsection{Ice formation, and O/C/N budgets}

In the previous sub-section, we described how H$_2$O, CO and N$_2$ become the most abundant volatile carriers of O, C and N, in molecular clouds, the building material of planetary systems. We now turn to the other important carriers O, C, and N, and how they together set the complete O, C and N budgets.  

In molecular clouds, we can observe the sequential build-up of icy grain mantles, which at the onset of star formation constitute the main reservoir of volatiles (except for H$_2$ and He). Observations of ices in clouds and cloud envelopes surrounding protostars have been previously reviewed by \citet{Gibb04}, \citet{Oberg11c}, and \citet{Aikawa12}, and more recently by \citet{Boogert15}. In addition formation of different ice constituents has been studied experimentally and theoretically. CH$_3$OH formation from CO was the first system to be investigated in detail \citep{Hiraoka98}, followed by H$_2$O from O, O$_2$ and O$_3$ \citep{Hiraoka02, Watanabe02}, and CO$_2$ from CO and OH \citep{Ioppolo08,Miyauchi08,Mokrane09,Dulieu10}. By now there are experimentally verified atom-addition pathways to all major ice constituents \citep{Ioppolo11}. Note that additional ice chemistry may be needed to explain specific aspects of the observed ice morphology, such as the small amount of CO observed to be mixed with water ice \citep{Mennella04}.

\begin{figure}[htbp]
\begin{center}
\includegraphics[width=0.7\textwidth]{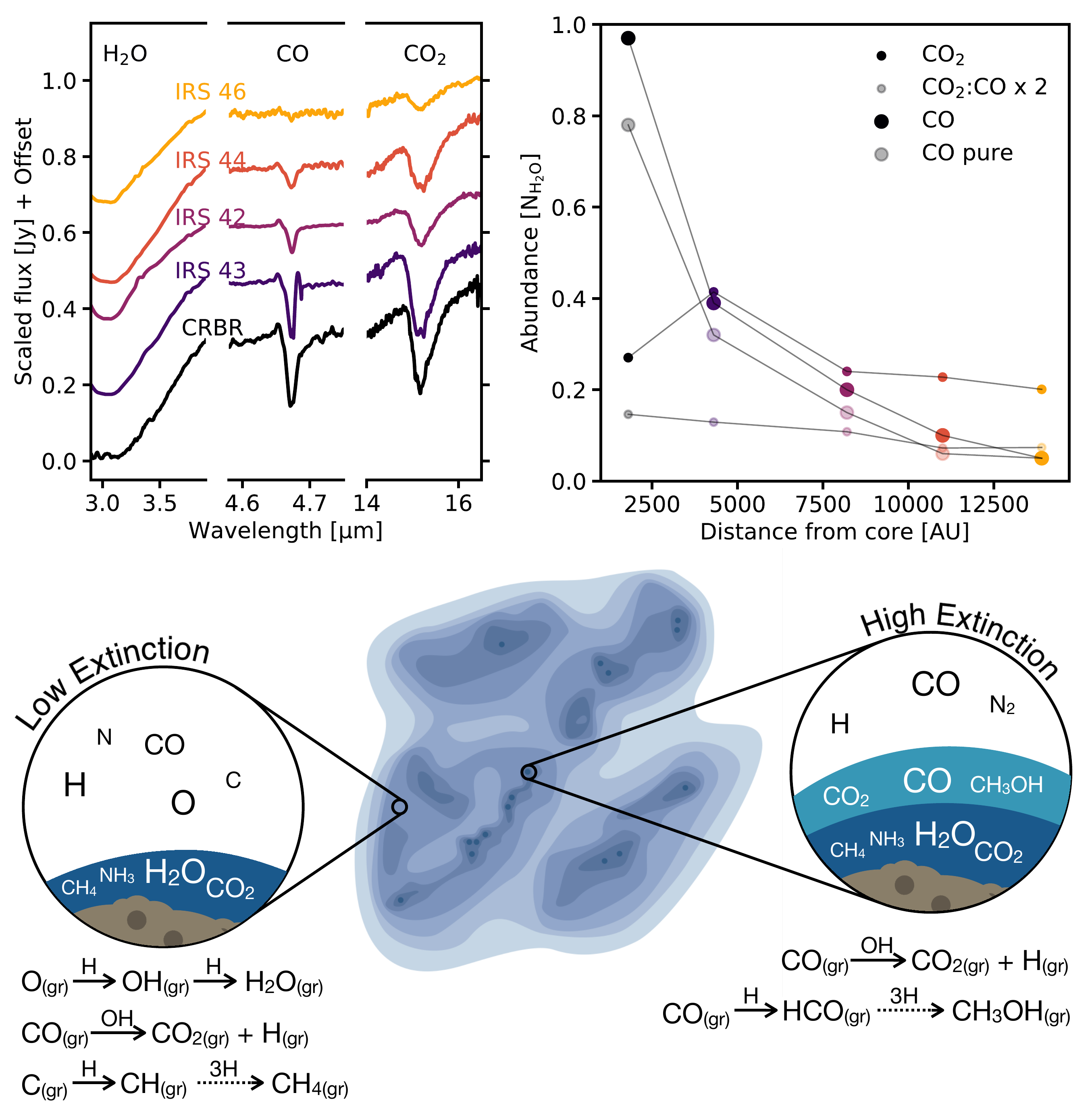}
\caption{Observations of sequential formation of ices.  ({\em Top:}) Ice spectra (top-left) and a radial profile of CO and CO$_2$ ices (top-right) obtained from spectra towards background sources that lie behind the $\rho$~Oph~F cloud core.  All ice abundances are relative to water ice and are given at the projected distance from the center of the core.  CO itself is listed as ``CO pure'' (CO in CO-dominated ice) and as ``CO'', which is the total CO ice content. Similarly we show the total CO$_2$ abundance, and the abundance of CO$_2$ in the CO-dominated ice layer.  ({\em Bottom:}) Schematic of ice layers and ice chemistry at Low Extinction where temperatures are warmer (T $>$ 10~K) and the UV radiation is diluted but present and High Extinction where the gas and dust are cold with UV radiation extinguished. Data from \citet{Pontoppidan06}. Image credit K. Peek.
}
\label{fig:ice-obs}
\end{center}
\end{figure}

Here be briefly review the observational and laboratory/theoretical results most pertinent for establishing the composition and morphology of typical icy grains during star and planet formation, and the relative importance of different volatile and non-volatile O, C, and N carriers. Figure \ref{fig:ice-obs} shows ice spectra observed towards protostars in $\rho$-Oph F, and how the ice composition changes with distance from the central dense core \citep{Pontoppidan08}. As established in the previous sub-section, water ice forms early, and therefore shows up throughout the cloud. The other two major ice constituents, CO and its derivative CO$_2$, increase in abundance with respect to H$_2$O towards the cloud core, increasing the C/O ratio in the ice. The rate of increase for CO and CO$_2$ are quite different, with CO increasing much more rapidly, which suggests that a substantial amount of CO$_2$ co-forms with the water ice. 

Based on these kinds of ice mapping observations as well as detailed characterizations of ice spectra \citep{Pontoppidan04,Pontoppidan08b,Boogert08}, the initial icy grain mantles are composed mainly of H$_2$O and CO$_2$. The early-forming water-dominated ice layer should also be the main host of CH$_4$ and NH$_3$, which are expected to form mainly from atomic carbon and nitrogen before N and C are completely converted into CO and N$_2$, respectively \citep{Tielens82,Garrod11}. The preference for CO$_2$ to co-exist with H$_2$O is explained by its  major formation pathway through CO+OH \citep{Ioppolo11b,Garrod11}. The OH forms in the ice through H addition to O, and can then either form H$_2$O through the addition of a second H, or CO$_2$ through a reaction with a CO molecule residing on the grain surface. The relative mixing ratio of H$_2$O and CO$_2$ in this ice layer will depend on the relative reaction rates between OH+H and OH+CO, which in its turn will depend on the gas-phase abundances of CO and H, as well as their different mobilities on grain surfaces.

Closer to the cloud core, elevated densities result in a CO freeze-out rate that is too high for H-atom activated grain-surface reactions to keep up, resulting a mainly CO-containing outer ice mantle. Depending on the exact conditions, some portion of the deposited CO does react with H atoms, produced through cosmic ray interactions with molecular hydrogen even in the densest part of the cloud, to form CH$_3$OH. Figure \ref{fig:ice-obs} (lower panels) illustrates this aspect of layered ice formation, while the upper panels show the rapid increase in CO-ice content deeper into the $\rho$~Oph~F core \citep{Pontoppidan06}. We expect that N$_2$ is either mixed in with the CO ice or frozen out on top since the N$_2$ freeze-out and sublimation characteristics closely resemble that of CO -- it is only about 10\% more volatile \citep{Oberg05,Bisschop06,Nguyen18}. The resulting ice morphology is important for the chemistry of planet formation for two reasons: 1) it regulates what kind of organic chemistry can take place in the ice since  not all ice reactants are mixed together, and 2) it allows for many of the hyper-volatiles residing in the CO-dominated ice, such as CO and N$_2$ to readily sublimate when the ice is heated, while other hyper-volatiles, such as CH$_4$ may be more difficult to dislodge from within the water ice lattice. The latter is key to estimate the O/C/N ratio of solids residing at different temperatures in protoplanetary disks.

\begin{figure}[htbp]
\begin{center}
\includegraphics[width=0.7\textwidth]{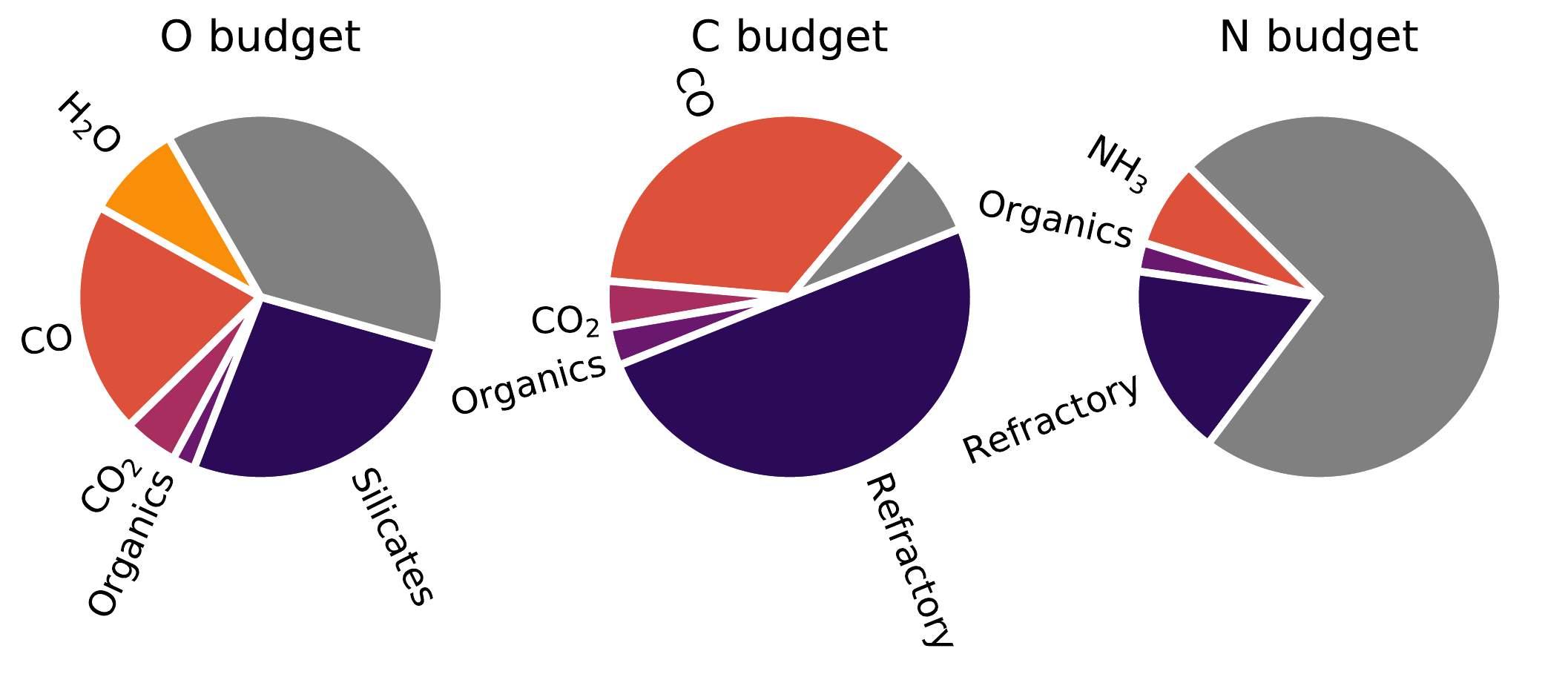}
\caption{The O, C, and N budgets at the onset of star and planet formation, based on observations of ice abundances \citep{Boogert15}, standard assumptions about CO, silicate and refractory carbon abundances, and solar system comet measurements of refractory nitrogen. Grey indicates an unknown component, which in the case of N is likely mainly N$_2$, while the nature of the missing O carrier is unclear.}
\label{fig:pie-chart}
\end{center}
\end{figure}

To date, icy mantle and gas-phase volatile compositions have been estimated in 10s of lines of sight. In Figure \ref{fig:pie-chart}, we combine these constraints with observations of solar system comets to estimate the O, C and N reservoirs at the onset of star formation. To start with, substantial parts of the O, C, and N budgets are refractory. In the case of O, this is mainly silicate grain cores, whose abundance can be estimated from observations \citep{Whittet10}, which constitutes about a quarter of the O budget assuming a Solar O abundance \citep{Asplund09}. Water, CO$_2$ and organic (CH$_3$OH and CH$_4$) ices make up another 13\%, and CO another 20\% (adopting a total CO gas + CO ice abundance of 10$^{-4}$ with respect to n$_{\rm H}$, the number of hydrogen nuclei). The remaining $\sim$40\% of oxygen is unaccounted for and this is a major mystery in interstellar studies \citep[see also,][]{Poteet15, Jones19}. There is some evidence for a hidden water ice reservoir;   in Orion, much of the oxygen not carried by CO, CO$_2$, and interstellar dust is found in water vapor \citep{Neill13}, which suggests that water ice is under-counted in interstellar ice studies, but this requires further investigation. For carbon, the accounting is somewhat easier as there are only four major carriers, CO, CO$_2$, organics, and refractory carbon. About half of the carbon is thought to be refractory based on observations of carbon depletion in the diffuse interstellar medium \citep{Mishra15}, which implies that carbon is almost equally split between refractory and volatile material. Of the volatiles, a majority is hypervolatile, i.e. CO. In the case of nitrogen, only a small part of the budget is directly observed in the form ammonia \citep{Boogert15}, and N-containing volatile organics \citep{Rice18}. Based on solar system comets, a substantial portion of N may be in more refractory organics and salts \citep{Altwegg20}. The majority of N is unobserved, and is likely present in the form of N$_2$.

\subsection{Building up the organic reservoir}

In addition to dividing up O, C, and N into its major volatile reservoirs, the molecular cloud phase is repsonsible for forming the first generation of organic molecules, which constitute the feedstock of organic chemistry during the later stages of star and planet formation. This generation zero of organics \citep{Herbst09} forms through a combination of gas-phase and ice-phase chemistry and is observed through ice infrared absorption spectroscopy, and through rotational spectroscopy of gas-phase molecules both at the UV-exposed edges of clouds, and deep into cloud cores.

On grain surfaces the two most common organic molecules are CH$_3$OH and CH$_4$, which form through H addition chemistry as outlined above. Observationally there is a clear connection between catastrophic CO freeze-out and the gas-phase appearance of CH$_3$OH in cloud cores \citep{Caselli99,Bizzocchi14}, which likely originates from low levels of non-thermal sublimation of CH$_3$OH ice formed from CO ice. CH$_3$OH and CH$_4$ are also abundant in comets \citet{Mumma11}, highlighting the persistence of the molecular cloud organic reservoir from cloud collapse to planetesimal formation. In addition to CH$_3$OH and CH$_4$, HNCO also seems to form through ice chemistry and is detected in the ice in its ionized form (OCN$^-$). There is, furthermore, observational evidence for that some of this initial organic ice reservoir is converted into more complex organics  in interstellar clouds \citep{Oberg10a,Bacmann12,Cernicharo12,Vastel14,Jimenez-Serra16,Scibelli20}. Different low-temperature ice pathways, and ice-dependent gas-chemistry pathways to complex organic molecules were reviewed in \S\ref{sec:Reactions}, and it is currently unclear which one is the most important for complex organic molecule formation in clouds. There is some evidence for a changing complex organic chemistry between cloud core and cloud edges, which suggests that different pathways may drive the growth of complexity at different stages of cloud and core evolution \citep{Jimenez-Serra16}.
 In either case, the conversion efficiency is expected to be low, and we expect that the organic ice reservoir in molecular clouds mainly consists of CH$_3$OH and CH$_4$.

Another pathway to organic molecules is through gas-phase ion-molecule and neutral-neutral reactions. Such pathways are responsible for small common organics like HCN, as well as for larger unsaturated organics such C$_4$H \citep{Herbst09}. With water vapor and molecular oxygen absent from the gas phase, the gas has C/O $\gtrsim$ 1 (with most of the gas-phase carbon and oxygen carried by CO, since water is frozen on grain surfaces), which is at least in part responsible for the oxygen-poor nature of these organics \citep[see, e.g.,][]{Langer84}. Gas phase chemistry is also implicated in the formation of benzene, benzene derivatives, and small polycyclic hydrocarbons in molecular clouds \citep{McGuire18a}. Compared to complex organic molecules formed in the ice, large organic molecules formed in the gas-phase tends to be oxygen and hydrogen poor. They therefore constitute an almost orthogonal reservoir of organic molecules that can become incorporated into planetesimals if they survive through cloud collapse and disk formation. Such survival is quite likely if the molecules freeze-out, which is expected in the dense parts of clouds where freeze-out time scales are short, and become embedded in the icy grain mantles.

Complex organic molecules are also detected in classical PDRs. The organic chemistry there seems to favor hydrocarbons and nitrogen-containing organics over oxygen-bearing molecules \citep{LeGal17}. It is currently unclear whether this is due to an efficient gas-phase organic chemistry operating under elevated C/O conditions, due to oxygen lock-up in water ice and /or photoinduced carbon-grain destruction enriching the gas in carbon, or whether we are there witnessing the products of an oxygen-starved grain-surface organic chemistry \citep{Guzman15b,LeGal19b}.  In either case, the organic composition found in PDRs demonstrates that there are more pathways to organic complexity in the molecular cloud stage than has traditionally been assumed. Thus even if the most abundant organics are small at the onset of star formation, there are already more complex organics present, and their nature will depend on the contribution from ice, gas, and PDR organic chemistry.

\subsection{Isotopic fractionation}

A final aspect of molecular cloud chemistry of importance to decode the chemical environment within which planets form is isotopic fractionation, and the resulting isotopic imprints in hydrogen, carbon, nitrogen, and oxygen. 
The isotopic composition of a molecule is set at its formation, and because most isotopic fractionation processes are very sensitive to either photon flux, or gas or grain temperatures, the D/H, $^{13}$C/$^{12}$C, $^{15}$N/$^{14}$N, and $^{18}$O/$^{16}$O ratios in molecules can be used to pinpoint when and where they formed. This is famously used to constrain the origin of the Earth's oceans. It is more generally, a tool to understand the origins of volatile reservoirs in the solar system, as well as in disks where exoplanetary systems are currently assembling. In this sub-section we review major isotopic fractionation pathways for H, C, N, an O in molecular clouds and how they are predicted and observed to affect the isotopic compositions of volatiles formed at this evolutionary stage. The review is by necessity brief, and for more detailed descriptions of isotopic fractionation chemistry in interstellar and circumstellar environments we refer the reader to \citet{Aikawa01}, \citet{Visser09}, \citet{Aikawa12}, \citet{Persson13}, \citet{Ceccarelli14}, \citet{Heays14}, \citet{Roueff15}, \citet{Furuya15}, and \citet{Furuya18}.

\subsubsection{Hydrogen}

\begin{figure}[htbp]
\begin{center}
\includegraphics[width=0.8\textwidth]{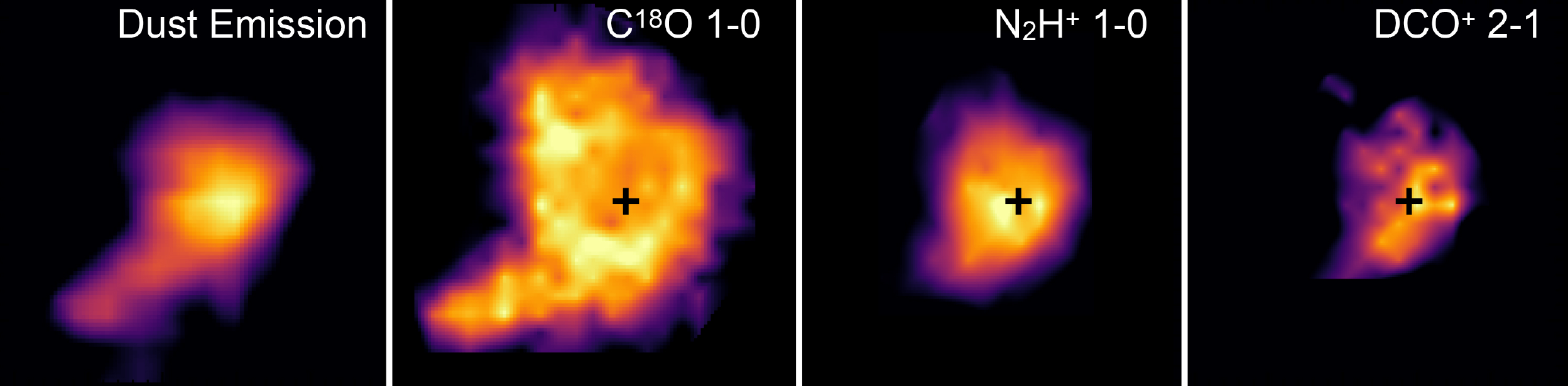}
\caption{Montage of mm-wave emission distributions of dust thermal continuum and molecular emission lines within the Barnard~68 prestellar core.  Dust emission is the 850~$\mu$m flux from \citet{Bianchi03}, C$^{18}$O 1--0 and N$_2$H$^+$ 1--0 from \citet{Bergin02}, and DCO$^{+}$ 2--1 from \citet{Maret07}.   The cross shown in molecular emission line images for C$^{18}$O, N$_2$H$^+$ 1--0, and DCO$^{+}$ 2--1  denotes the location of the dust continuum flux peak or the center of the core traced by the maximum of the dust column.}
\label{fig:b68}
\end{center}
\end{figure}

Deuterium is formed as part of big bang nucleosynthesis \citep{Boesgaard85, Galli13} with an abundance of D/H $\sim 2.5\pm 0.2 \times 10^{-5}$. 
Deuterium fractionation occurs  at cold  temperatures due to zero-point energy differences between hydrogenated and deuterated ions;   
 because of its larger mass, D forms slightly stronger bonds than H, which makes it favorable to transfer D from HD into ions and molecules with strong bond interactions. The primary reactions within the deuterium fractionation chemical networks are \citep{Millar89, Roueff15}:

\begin{equation}
    {\rm H_3^+ + HD \leftrightarrow H_2D^+ + H_2 + 232\,K},
    \end{equation}
    
    
\begin{equation}
    {\rm CH_3^+ + HD \leftrightarrow CH_2D^+ + H_2 + 654\,K}.
\end{equation}

If the temperature is cold ($<$~30~K for H$_3^+$ and $<$~300~K for CH$_3^+$) the forward reaction is favored and the back channel is inactive.  This is slightly more complicated as the endothermicity of the back reaction depends on the spin states of the products in particular that of H$_2$ \citep{PineaudesForets91, Pagani92, Flower06,Sipila15}. Once H$_2$D$^+$ or CH$_2$D$^+$ has formed they can efficiently transfer a deuteron to a range of other molecules, e.g. to CO to form DCO$^+$ \citep[e.g.,][]{Turner01}. Some fractionation could thus be expected throughout cloud regions that are $<$30--300~K. Observationally, large abundances of deuterated molecules like DCO$^+$ are only observed in the coldest, densest, and most well shielded parts of molecular clouds (Figure \ref{fig:b68}), i.e. in molecular cloud cores, which suggests that something in addition to low temperatures is needed to substantially deuterate molecules. This something is CO depletion. Figure \ref{fig:b68} shows that the dense core, here traced by dust emission, is associated with depletion of CO gas (as traced by its rare C$^{18}$O isotopologue), and excess DCO$^+$. The reason CO depletion is needed for efficient deuterium fractionation, is that the chemistry depends directly or indirectly on the abundance of the H$_3^+$ ion, and this ion is rapidly destroyed by gas-phase CO (H$_3^+$ $+$ CO $\rightarrow$ HCO$^+$ + H$_2$)  \citep{Lepp87,Caselli02}. 

\begin{figure}[htbp]
\begin{center}
\includegraphics[width=0.6\textwidth]{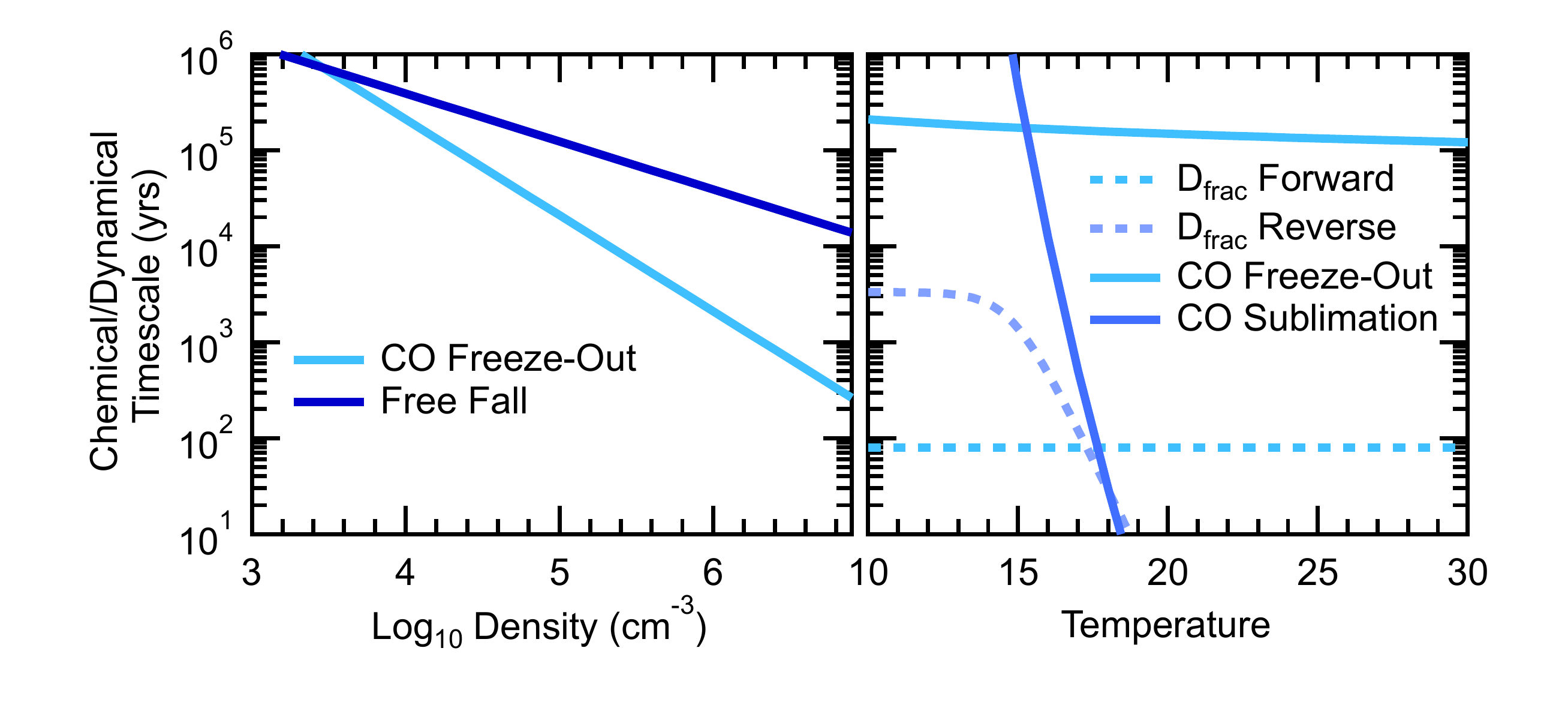}
\caption{Chemical and dynamical timescales as a function of density (assuming T$_{gas}$ = T$_{dust}$ = 10 K), and temperature (assuming n$_{\rm H_2} = 10^4$~cm$^{-3}$).  CO freezeout using formalism of \citet{Hollenbach09} and sublimation using the CO binding energy measured by \citet{Fayolle14}.  For deuterium fractionation the timescales refer to the forward and reverse of the following reaction:
H$_3^+$ $+$ HD $\leftrightarrow$ H$_2$D$^+$ $+$ H$_2$ $+$ 232~K.
We use the rates given by \citet{Hugo09} and assume that the o/p ratio of the reactants and products are set by equilibrium at the specific temperature using the approximations described by \citet{Lee15}.  Under these assumptions the reverse reaction begins to activate near 17~K, but this transition temperature will depend sensitively on the history of ortho and para H$_2$ during cloud and core formation \citep{Flower06,Pagani13, Furuya15}.}
\label{fig:time}
\end{center}
\end{figure}

CO depletes from the gas-phase through freeze-out onto grains in cloud cores when grain temperatures are $<$18~K \citep{Bisschop06} if CO mainly binds to other CO molecules, and at somewhat higher grain temperatures if CO freezes out on top of water ice \citep{Collings03,Fayolle16}. Cloud cores are colder than this, and the level of depletion in a given core therefore depends on the depletion time scale, which is a direct function of core density \citep{Bergin95,Caselli99}. Figure \ref{fig:time} shows that for typical core densities the freeze-out time is less than 10$^5$ years. This is smaller than the free-fall time scale and we can therefore expect CO to be frozen out and an efficient deuterium chemistry activated by the time a cloud core collapses to form a star and planetary system. This is further illustrated in the right-hand panel of Fig.~\ref{fig:time}, which shows that for temperatures $<$17~K, the forward deuterium fractionation reaction pushes the D into H$_2$D$+$ such that H$_2$D$^+$/H$_3^+$ $\gg$ HD/H$_2$, often by orders of magnitude \citep{Millar89}. 

For the chemistry of planet formation, the most important lasting  effect of this fractionation process is a by-product; elevated H$_2$D$^+$/H$_3^+$ result in an excess of atomic deuterium in the gas (D/H $\gg$ HD/H$_2$) because in dense cloud cores D and H atoms are produced through recombinations between deuterated and  protonated molecules and electrons. Excess H$_2$D$^+$ therefore naturally produces excess D atoms. These D atoms collide with grain surfaces and participate in chemical reactions that lead to D enrichments in the molecular ices for both water and organics \citep{Tielens83, Taquet13}. All ices discussed in the previous sub-section that form through hydrogen addition will be fractionated in D if they form in a cloud (core) region where D/H $>$ HD/H$_2$. 

Based on observations and theory, different ice components form at different times, and therefore at different cloud temperatures and densities, which should result in different levels of deuterium fractionation \citep{Caselli12}. Most water forms prior to core formation and should therefore be less deuterated than molecules like CH$_3$OH, which forms in the dense cloud core. Still, substantial H$_2$O deuterium fractionation is expected \citep{Furuya15,Furuya16}, which explains why all water in the solar system is enriched in water \citep[e.g.,][]{Cleeves14}. CH$_3$OH and organics forming from CH$_3$OH should have higher levels of deuteration than H$_2$O, and this is indeed observed \citep{Parise02,Coutens12,Persson14,Taquet19}. Whether these high D/H levels in organics are preserved during planet formation is less clear than in the case of water, and is the subject of ongoing studies. Elevated D/H ratios in organics in comets and carbonaceous chondrites compared to water \citep{Alexander07,Alexander10,Altwegg19} suggest that the interstellar D/H patterns persist throughout planet formation, but disk chemistry models also show that a disk {\it in situ} origin of the observed organic deuterium fractionation cannot be excluded \citep{Cleeves16}.

\subsubsection{Carbon, Nitrogen, and Oxygen}

The chemical fractionation via ion-molecule reactions with low endothermicity discussed above for H is also present for C, and N, and perhaps also for O; a similar zero-point energy difference exist between the heavy isotopologues containing $^{13}$C, $^{18}$O, and $^{15}$N, and the normal isotopologues, but the relative mass differences are smaller and the expected fractionations are therefore less.   For carbon, a central reaction in the ion-molecule chain that powers fractionation is $^{13}$C$^+$ $+$ $^{12}$CO $\leftrightarrow ^{12}$C$^+$ $+$ $^{13}$CO $+$ 35~K; this does lead to a slight excess of $^{13}$CO in cold regions \citep{Langer90, Milam15}. 
A similar set of reactions are found in the nitrogen pool that again favor heavy element enrichment at low temperature \citep{Terzieva00}; however, the overall efficacy of various pathways are currently a matter of debate \citep{Roueff15, Wirstrom18}.  
For oxygen, \citet{Langer89} demonstrated that there is little ion-molecule powered fractionation in major oxygen carriers such as CO, while \citet{Loison19} finds that there may be effect  for other, less abundant species.

For C, N and O there also exists another, often more important isotope fractionation pathway: 
isotopic selective photodissociation.  Both CO (main carrier of C and O) and N$_2$ (main carrier of N) are dissociated via molecular bands or lines in discrete regions of the UV spectrum as opposed to continuous photodissociation cross-sections from 912 to $\sim$2000~\AA .  These spectral lines can saturate ($\tau \gg 1$) and any molecules downstream from the source of radiation will remain safely shielded from molecule-destroying UV radiation.  As the radiation transfer is limited by absorption via electronic-state transitions, the optical depth of a given line, $\tau$, has a direct dependence on the molecular abundance.  Lesser abundant isotopologues ($^{13}$CO, C$^{18}$O, C$^{17}$O, $^{15}$NN) have lower line optical depth and, hence, the onset of self-shielding is found in successively deeper layers than more abundant isotopologues.  As a result, there will a deficit in the heavy, less abundant molecular isotopologues, and an excess of the heavier (less abundant) elemental atoms in the outer surface layers of clouds. For example, for oxygen, $^{12}$C$^{16}$O self-shields prior to the onset of $^{12}$C$^{18}$O self-shielding.  In gas layers that lie between these two self-shielding transitions there is an excess of $^{18}$O.  Similar effects are seen for nitrogen \citep{Heays14, Visser18,Furuya18} and carbon \citep{Rollig13}.  

This isotopic stratification will affect the isotopic patterns of ices formed from atomic O, C and N. In particular, water ice that forms early and therefore in UV exposed regions of molecular clouds should be enriched in $^{18}$O because of isotope specific photodissociation. If these ices survive the star and planetesimal formation process, we would expect to see $^{18}$O enrichments in bodies that formed volatile-rich \citep{Yurimoto04, Lee08}.  Indeed isotopic selective photodissociation is a prominent theory for the origin of heavy element enrichments detected in meteorites for both O and N \citep{Clayton93, Marty12}.  Such an enrichment could also be due to isotope-specific photodissociation in planet-forming disks, since similar to clouds, disks too have a UV exposed PDR layer where this process should be important \citep{Lyons05, Visser18}, and it is currently unclear whether this isotopic signature is a sign of volatile inheritance or not.

\subsection{Summary: Setting the Chemical Trajectory}
 
At the onset of cloud formation, the chemistry is constrained by the locking up of $\sim$1/4 of the oxygen and $\sim$1/2 of the carbon in refractory material. 
During cloud formation, the remaining carbon, oxygen and nitrogen become mainly incorporated into CO, water and N$_2$, respectively. The prominence of these C, N, and O carriers appear universal across star forming regions, and across time during star and planet formation.
Since water ice, and the unidentified oxygen carrier(s), are significantly less volatile that CO or N$_2$ this means that for much of the cloud-chemical evolution, the gas-phase elemental budgets have C/O $\sim$ 1, and N/O $\sim$ 0.2 (i.e. excess C and N with respect to O). Though many details remain uncertain about this chemical evolution, we can say with confidence that at the beginning of cloud collapse to form a star and planetary system, the carriers of the C, N, and O elemental pools are already established.

In addition to establishing the main volatile carriers, dense cloud chemistry is responsible for producing the organic feedstock molecules that drive a more complex organic chemistry at later evolutionary stages. Ammonia, methane, and methanol all form on the surfaces of interstellar grains and are incorporated into the icy grain mantles. Finally clouds and cloud cores are seats of efficient isotopic fractionation. At cloud edges, CO and N$_2$  self-shielding results in a build-up of the main isotopologue and a relative depletion in the minor isotopologues. In the cold cloud cores heavier isotopologues are instead favored due to small differences in zero-point energies between normal and main isotopologues. This is especially important for deuterium fractionation, which leaves imprints on organics and water that can still be seen in solar system volatiles today.

\section{Chemical inheritance and transformation during the protostellar stage}

Protostars form through the collapse of cloud cores. In our understanding of the chemistry of planet formation, this stage plays a key role because, first, it is during the warm-up of infalling cloud material towards the central protostar that much of the icy molecular cloud chemistry described in the previous section is revealed. Second, warm-up of interstellar grains in the protostellar envelope activates new chemical pathways that changes the compositions of the future solid building blocks of planetesimals and planets. Third, the protostellar disk that forms at this stage is the precursor to the planet-forming disks treated in the next section, and the balance between preservation or inheritance and reset at this stage provides the initial chemical conditions for planet formation. In this section we review the protostellar organic chemistry, and the chemistry or protostellar disks, after a brief review of the chemical structures of low-mass protostars (analogs to the protosun) and their surrounding environment. 

\subsection{The chemical structure of Solar-type protostars \label{sec:protostar}}

\begin{figure}[htbp]
\begin{center}
\includegraphics[width=0.8\textwidth]{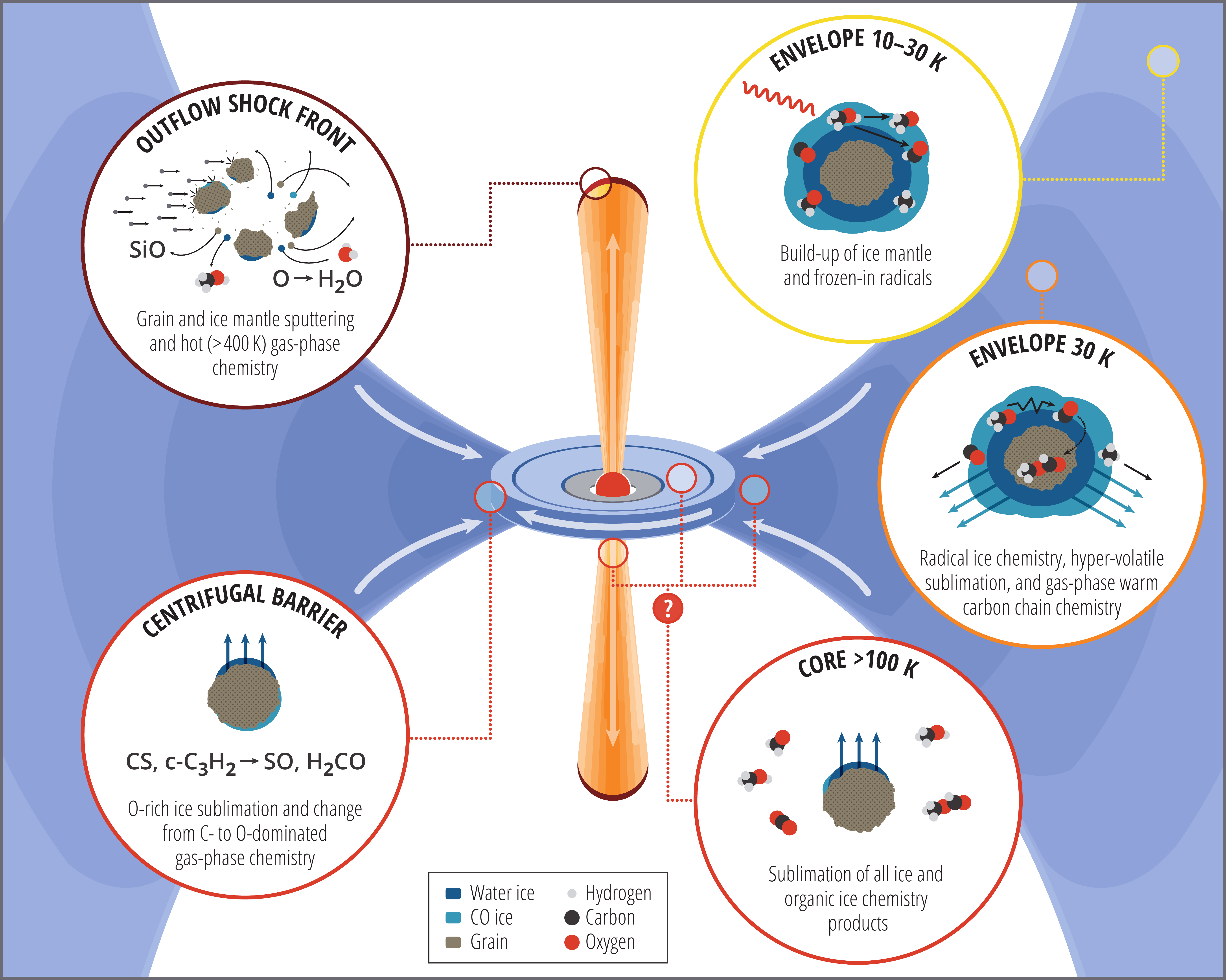}
\caption{The main structural elements of a protostar are the core, envelope, disk, and outflow. The envelope presents a radial temperature gradient, which results in a radially dependent gas and grain surface chemistry regulated by sublimation, radical ice chemistry, and temperature-dependent gas chemistry. The disk and the disk-envelope interface presents its own set of specific chemical conditions, including weak accretion shocks. Along the outflow we expect to see both shock and UV driven chemistry. Image credit: K. Peek}
\label{fig:ps-cartoon}
\end{center}
\end{figure}

Low-mass or Solar-type protostars ($<$ 2M$\rm_\odot$, from now on simply referred to as protostars) present multiple distinct chemical environments, which are not present during the earlier stages of star formation \citep{Caselli12}. Figure \ref{fig:ps-cartoon} illustrates the rather complex and dynamical structure characteristic of protostars. The central protostar is surrounded by a large envelope of infalling cloud material, which has an inward temperature gradient due to heating from the central protostar. At large distances, 1000s of AU, the gas and dust temperatures are low ($<$20~K), and the chemistry is initially a continuation of the processes found in molecular cloud cores. As gas and dust warm up above 20--30~K, multiple chemical changes occur at once. First the ice-gas equilibrium for the most volatile species such as CH$_4$, CO and N$_2$ changes, resulting in substantial sublimation of hypervolatiles not trapped in the water ice mantle. This changes the availability of reactants in the gas-phase, and especially the sublimation of CH$_4$ at these intermediate distances has been proposed to drive a warm carbon-chain chemistry \citep{Sakai08}. Second, at 20--30~K, some molecules and radicals begin to diffuse in ice mantles, resulting in a wave of ice organic chemistry \citep{Garrod08}.

Closer to the protostar, grains heat up above 100~K resulting in complete ice sublimation and the seeding of the gas-phase with water and a range of ice-produced organic molecules, forming a so called hot corino\footnote{Chemically rich ``hot cores'' around massive protostars have been known for a long time. The name ``hot corino'' was introduced by \citet{Bottinelli04a} to classify the low-mass or protosun analog of a hot core.}, which may then chemically evolve further \citep{Herbst09, Caselli12}. At similar scales as the hot corino we expect, and sometimes observe, the formation of a protostellar disk \citep[e.g.,][]{Tobin12}. The envelope-disk interface can result in mild shocks and therefore shock-driven sublimation of ices, again seeding the gas with ice-chemistry products, and gas-phase chemistry \citep{Aota15, Miura17}. In the disk itself, densities are high compared to cloud and envelope densities, enabling new kinds of gas-phase chemistry, possibly even 3-body reactions close to the protostar.

Finally, protostars generate outflows at a range of velocities, and the interaction of these outflows with the remnant envelope produces shocks that sublimate icy grain mantles, sputter the grain core, enables hot gas phase chemistry, and produces UV fronts that result in PDR like chemistry along the outflow cavity walls \citep[e.g.][]{Arce08}.\footnote{There are several seminal references regarding shock physics  and the reader is referred to these for more information  \citep{Draine79, Hollenbach79, Flower85, Hollenbach89, Neufeld89, Kaufman96}.} The prevalence of water vapor in outflowing gas is believed be partially induced by the activation of neutral-neutral reactions with activation barriers that when overcome lead to rapid water production \citep{Wagner87, Kaufman95}. A similar chemistry regulates the water vapor abundance in the hot gas close to the star in protoplanetary disks \citep{Pontoppidan10, Bethell11}.

\subsection{An increasing organic complexity}
\label{sec:protostar-organics}

Molecular cloud and cloud core chemistry is responsible for the production of organic feedstock molecules during star and planet formation, while protostellar envelopes are  organic chemistry factories that transform a large fraction of this feedstock into more complex, but still fairly volatile, complex organic molecules \citep{Herbst09}. This transformation is the subject of this sub-section. Before reviewing this process, it is important to remind ourselves that there is mounting evidence for a limited complex organic chemistry in cloud cores and at the UV-exposed edges of clouds. The importance of this chemistry for the organic inventory during planet formation is not yet clear. The current mainstream view, which is what is described here, is that it is minor compared to the organic chemistry activated during protostellar formation. This view may change, however. Observations of COMs in outflow shocks, which may originate from sublimation of interstellar ices \citep{Arce08}, and  of PAH formation in clouds \citep{McGuire18a} suggest that substantial amounts of complex organics may form in clouds prior to the onset of star formation. There are also theoretical models that suggest that complex molecules can form in icy grain mantles at very low temperatures \citep[e.g.][]{Shingledecker18}, and the reader is encouraged to closely monitor further developments.

The chemical transformation associated with protostars begins at the early stages of protostellar formation when icy grains streaming towards the contracting and increasingly hot core begin to heat up. The first important temperature threshold is around 25-30~K. At these temperatures, the most volatile ice constituents CO, N$_2$ and CH$_4$ begin to sublimate. The rapid increase of CO around protostars due to this process has been observationally confirmed 
\citep{Jorgensen02}. The release of CH$_4$ into the gas-phase has been proposed by \citet{Sakai13} to fuel an efficient warm carbon chain chemistry resulting in high abundances of e.g. C$_4$H and other unsaturated hydrocarbons in the envelopes of protostars \citep{Law18}. These carbon chains have a lower volatility than CH$_4$ and may freeze out on the grains increasing its inventory of unsaturated organics. There is evidence for that some such unsaturated organics, whether they originate from the molecular cloud or the protostellar envelope, survive incorporation into the planet-forming disk and further into planetesimals, based on their presence in comets \citep{Altwegg20}.

Around 20-30~K, there are also new chemical pathways activated in the icy grain mantles, which enable the transformation of organic ices produced in earlier phases (e.g. CH$_4$ and CH$_3$OH) into a wide range of larger and more complex organic material. The onset of this second generation of ice chemistry is powered by two factors: the build-up of radicals in cold ices, and diffusion of these radicals at slightly elevated ice temperatures \citep{Garrod06,Garrod08,Garrod13}. Radicals form in ices through incomplete hydrogenation \citep{Chuang16,Chuang17}, and through energetic processing by photons, electrons of cosmic rays of pre-existing ices \citep{Greenberg83, Oberg17}. Cosmic rays are present throughout molecular clouds, and in addition to their direct interaction with icy grains, they also produce secondary electrons, and a dilute UV field  \citep{Prasad83}. Energetic processing of especially CH$_3$OH ice is expected to be highly efficient at producing organic radicals such as CH$_3$, CH$_3$O and CH$_2$OH. At 10~K these are frozen into the ice. In the protostellar phase, initial grain heating to 20-30~K enables radical diffusion, resulting in a productive radical-radical chemistry, where e.g. ethanol forms from CH$_3$+CH$_2$OH \citep{Bennett07a,Oberg09d}.

Most ice-produced COMs are expected to survive until the bulk of ice sublimates around 100~K, 10-100~AU away from the central protostar.  Models implementing such a scheme have been successful at explaining complex organic abundances following sublimation (described next), but it is important to note that we are not able to observe the complex organics as they form in the ice. Beyond CH$_3$OH, CH$_4$, HNCO, H$_2$CO and HCOOH, organics have abundances below the $\sim 1$\% (relative to H$_2$O) threshold for detection via current infrared absorption studies \citep{Boogert15}, and are thus hidden from view until they sublimate and become possible to detect via gas phase spectroscopic techniques \citep{Widicus-Weaver19}. Perhaps our best view of pristine icy organics that have not experienced temperatures beyond 30~K will come from in-depth observations of protostellar outflow shock fronts. Some outflow shocks are associated with complex molecule detection and the interpretation is that these originate from rapidly sublimating ices as a protostellar outflow slams into the protostellar envelope \citep{Arce08,Lefloch17}. We note that there are ongoing programs such as SOLIS (Seeds of Life In Space) to map out how the ice organic chemistry revealed through outflow shocks may differ from the organic ice chemistry revealed in hot corinos \citep{Ceccarelli17}.

\begin{figure}[htbp]
\begin{center}
\includegraphics[width=0.6\textwidth]{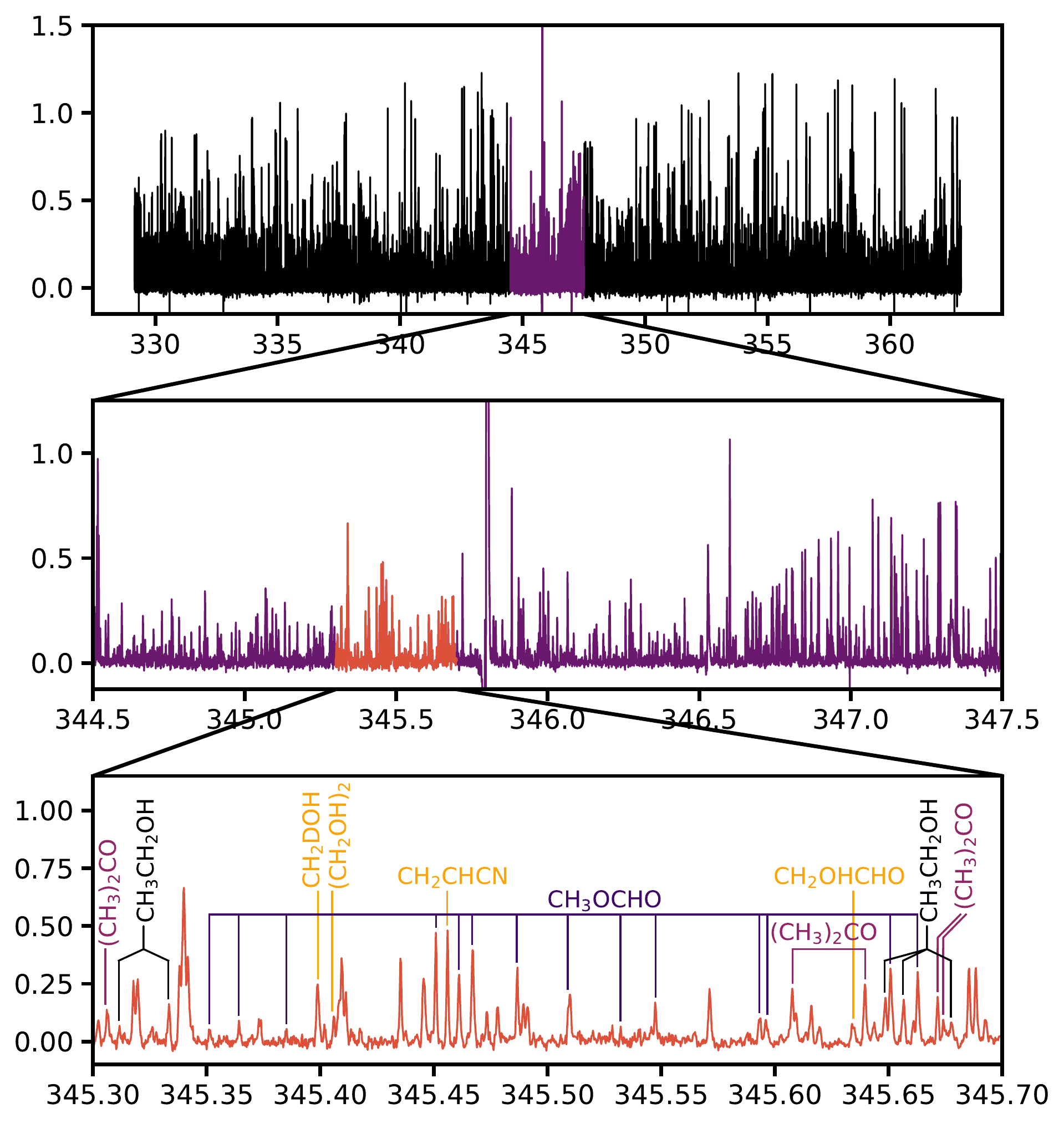}
\caption{40 GHz of the spectrum towards the Solar-mass protostars IRAS 16293-2422 from \citet{Jorgensen16}. The spectrum is acquired with ALMA at the organic 'sweet spot' next to source A in this binary star, approximately 50~AU from the source center. IRAS 16293 is one of a growing number of protostars that display a rich complex organic spectrum, which can be traced back to a hot molecular core in the vicinity of the protostar, where rapid ice sublimation enriches the gas-phase with complex ice chemistry products, as well as feeds a hot gas-phase organic chemistry.}
\label{fig:i16293}
\end{center}
\end{figure}

As icy grains continue to flow towards the central protostar they continue to warm up. When they reach 100--200~K, the bulk water ice sublimates, and most of the organic ice will co-desorb resulting in a hot, organic-rich core, i.e. a hot corino, since we are here interested in a Solar-like protostar. Hot corinos are observationally characterized by dense spectra at millimeter wavelengths due to the large abundances of warm, mid-sized organic molecules, which each has a large partition function and therefore a multitude of rotational transitions. The most well-known example of such a protostellar hot corino is IRAS 16293-2422, a protostellar multiple system, whose mm spectrum is shown in Fig.~\ref{fig:i16293}.  The spectrum is incredibly rich at mm, sub-mm, and far-infrared wavelengths, with most of the detected lines associated with organics and water \citep{vanDishoeck95,Zernickel12, Crockett14,Jorgensen16}. 

Over $\sim$40 organic molecules with 8 atoms or more have been detected in hot corinos, revealing a rapid growth in organic complexity in the evolutionary stage immediately preceding the formation of planet forming disks \citep{McGuire18b}. Some of the organics may originate from hot gas chemistry close to the protostar \citep{Charnley97}, but a majority of the inventory can be traced back to ice chemistry. Both water \citep{Jacq88, Persson13} and complex organics  \citep{Charnely97, Belloche16} in hot corinos are highly deuterated, which, as discussed in the previous section, can be traced back to the formation conditions of water and CH$_3$OH in molecular clouds and cloud cores \citep{Aikawa12}. If the hot corino molecules were mainly produced through gas phase chemistry, the chemical kinetics would rapidly push the D/H ratio back to $\sim 10^{-5}$ in $\sim$10$^4$ years \citep{Rodgers96}.  

One new aspect revealed by high-resolution images down to tens-of-au scales by ALMA, regards the geometry of the hot corino itself.  In general, the organics are assumed to lie in an evaporation zone extending into the heated envelope surrounding a young hot protostellar disk ( Fig.~\ref{fig:ps-cartoon}). In Figure~\ref{fig:PSD}, right panel, we show that protostellar disks may be important contributors to the complex organic emission at these scales. In the displayed source, complex organic molecular emission is visibly associated with the surface of a disk rather than the surrounding envelope. Methanol (CH$_3$OH) and more complex organic molecules, including ethanol,  CH$_3$OCH$_3$ and HCOOCH$_3$, have been detected towards a number of protostars with confirmed disks or disk candidates \citep{Sakai14a,Lee17,Codella18, Lee19, vantHoff18,Martin-Domenech19,Bergner19a}. These are the same kind of molecules that are signatures of hot corinos \citep{Herbst09}. Furthermore, the organic molecules in these disk sources are generally observed to be warm or hot, fulfilling commonly used criteria for hot corino emission. Whether there is a general causal connection between protostellar disk formation and hot corinos is currently unclear. 

In summary, protostellar envelopes are sites of chemical transformation of initially simpler organic molecules into more complex ones. The chemistry mainly takes place in icy grain mantles, but is only observable following ice sublimation close to the protostar, or in outflow shocks. The icy grains that become incorporated into  protostellar disks, pre-cursors of planet-forming disks, should be expected to contain a rich organic inventory, with organic molecules ranging in complexity from methanol to volatile complex organic molecules, and further to more refractory organics. Understanding this organic inventory is key to map out the initial organic chemistry on planets. We note, however, that despite the impressive spectra produced by mid-sized complex organic molecules, these molecules do not constitute a major elemental reservoir. Accounting of organic carriers of C and N within the complete spectrum of the best studied hot core in Orion \citep{Crockett14} suggests that volatile and semi-volatile organics carry only $\sim$1\% of interstellar carbon \citep{Bergin14}, a $\sim$few percent of total nitrogen \citep{Rice18}, and less than a percent of the oxygen. The main elemental carriers at this stage remain the same as in the molecular cloud and cloud core (e.g. H$_2$O, N$_2$, CO, CO$_2$).

\subsection{Protostellar disk chemistry: Reset and Inheritance}
\label{sec:reset}

Disks form during star formation to preserve angular momentum during the collapse and accretion of interstellar cloud material onto the central protostar \citep[e.g.][]{Tereby84}.
The initial stages of disk assembly close to the protostar are expected to be hot, and may result in complete atomization, removing any recollection of the chemistry of previous stages. As this material  viscously spreads outward, it will seed the outer and cooler disk regions with the resulting chemical products, which may be quite different compared to the interstellar inventory \citep{LyndenBell74, Cuzzi93, Ciesla10}.  Throughout disk formation,  infalling material is further subject to irradiation \citep{Visser09} and accretion shocks \citep{Neufeld94, Visser10, Aota15, Miura17}, which may also alter the chemical composition of material that accretes during later, less energetic phases, see, e.g. \citet{Lunine91}. However, as disk formation proceeds, the centrifugal radius where infall is being supplied to the disk rapidly moves outwards \citep{Shu77}. Thus there is substantial material supplied at large distances where any chemical reprocessing of supplied material will be minimal.  The central question then becomes whether the disk chemical composition is best described by chemical reset or by inheritance from the interstellar medium and protostellar envelope.

The solar system record provides evidence for both a chemical reset, and substantial inheritance. The volatility trend seen in elements in certain meteoritic classes \citep{Davis06} and in the Earth's lithosphere, where there is a correlation between the elemental abundance (normalized to Mg and CI chondrite) to the ``half-mass condensation temperature''\footnote{The half-mass condensation temperature is defined through a sequence of equilibrium condensation for a given element assuming an atomic gas of solar composition that is cooling from temperatures $>$2000~K.  For example, from \citet{Lodders03}, the majority of Mg removal from the gas coincides with the condensation of forsterite (Mg$_2$SiO$_4$) and enstatite (MgSiO$_3$) and 1/2 of the mass of Mg is removed from the gas and placed into a solid mineral at 1336~K. }  \citep[][]{McDonough95, Wood19}, suggest a complete ablation of ices and refractory grains at temperatures above 2000~K, followed by condensation of minerals through a sequence defined by equilibrium chemistry \citep{Grossman72}.
On the other hand, {\em all} solar system water carries a deuterium enrichment. This requires chemistry active below 30 K \citep[][]{Millar89} and cold storage at temperatures below the water ice sublimation temperature to maintain the enrichment. Efficient low-temperature production of H$_2$O is only possible in molecular clouds, which strongly indicates that most of the solar system's water was inherited \citep{Cleeves14}.  In addition, the volatile organic inventories in Comet 67/P and the protostar IRAS 16293-2422 are remarkably similar, further suggesting that much of the solar system organics were inherited as well \citep{Drozdovskaya19}. Neither a complete reset nor a complete inheritance scenario can explain the full solar system record and the only remaining answer is that both must have played a role during formation of the protosolar disk \citep[e.g.][]{Ciesla10}.  

\begin{figure}[htbp]
\begin{center}
\includegraphics[width=.5\textwidth]{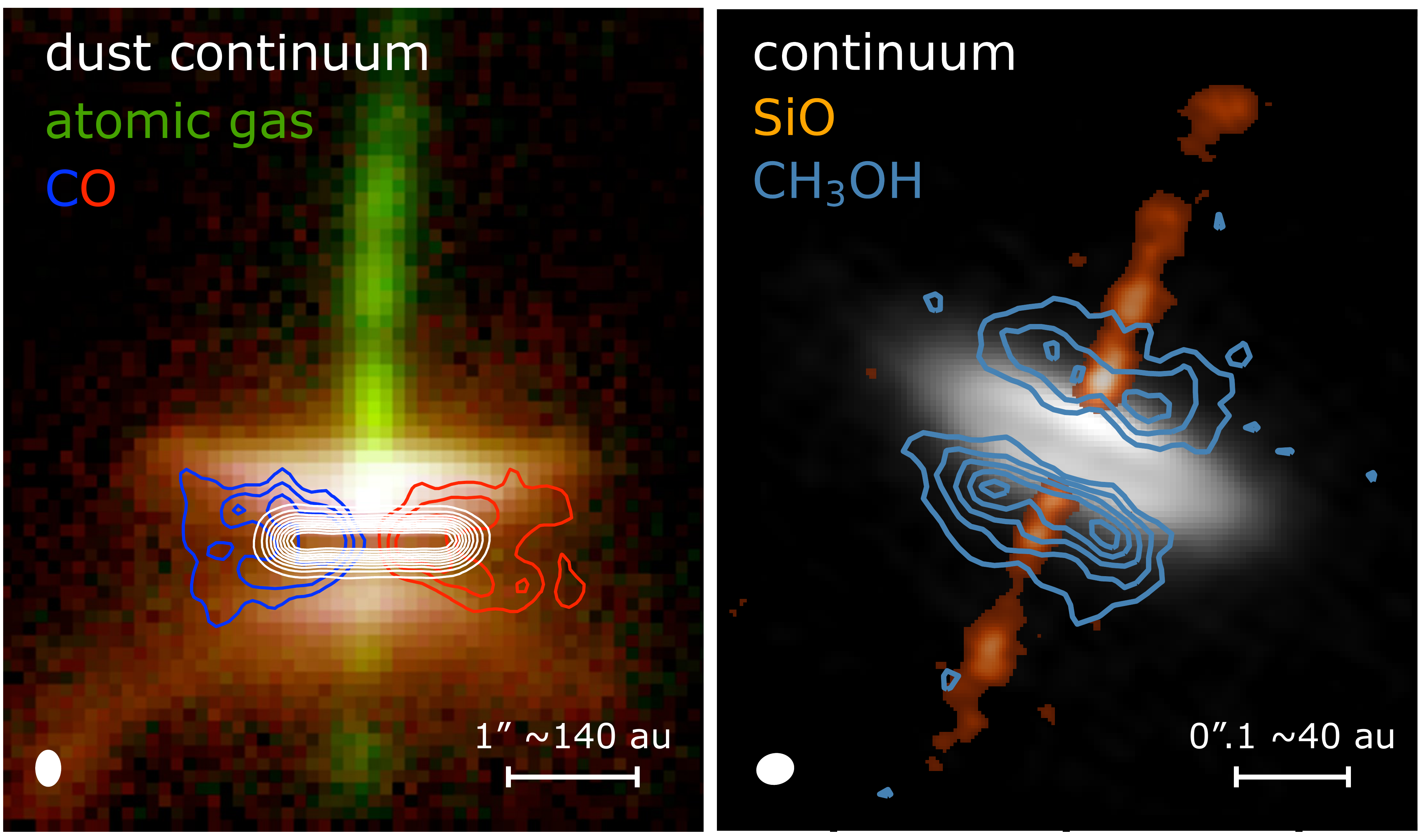}
\caption{{\it Left:} HST image of HH30 (credit: Chris Burrows (STScI), the WFPC2 Science Team and NASA/ESA) overlaid with ALMA observations of dust continuum (white contours) and blue and red-shifted CO 2--1 emission from \citet{Louvet18}, revealing a molecular disk. {\it Right:} ALMA observations of dust continuum from the HH 212 protostellar disk (white),  SiO emission tracing the protostellar jet, and CH$_3$OH emission, which appears to originate from the disk surface \citep{Lee17,Lee19}. In both panels the synthesized beams for the continuum are shown in the bottom left corner -- the molecular lines were observed at a comparable spatial resolution.}
\label{fig:PSD}
\end{center}
\end{figure}

This mixed model is also supported by observations of protostellar disks.
Figure \ref{fig:PSD} shows two recent spectacular ALMA images of molecular emission emerging from edge-on protostellar disks around precursors to Sun-like stars in nearby star forming regions \citep{Lee17,Lee19,Louvet18}. Towards HH30 (left panel), CO 2--1 emission is clearly observed to be coincident with the dust disk.
Since CO is main reservoir of carbon these and similar observations enable us to address whether a substantial amount of CO has been converted into other molecules during the protostellar disk phase, or whether this interstellar reservoir of carbon is preserved intact. \citet{Zhang20} finds that CO is preserved, while \citet{Anderl16} and \citet{Bergner20} find some evidence for CO depletion in evolved protostellar disks/sources, indicative of that while CO is being incorporated intact into the disk, there are some disk processes that over time convert CO into other species, such as CO$_2$ and organics, potentially changing the balance between different C and O carriers \citep{Bergin14, Reboussin15, Schwarz16, Eistrup18, Schwarz18, Bosman18}.

The hot ($>$~1000~K) disk phase inferred from the inner ($<$2.5~au) solar system volatility trend has not been directly observed \citep{Persson16, Aso17, vantHoff17, delaVillarmois19, vantHoff20}, which suggests that a ``hot" reset phase must occur fairly early ($\sim$first 100,000 years) during disk formation. By the time protostellar disks are observable they appear too cold even for substantial water vapor \citep{Harsono20}, implying that water is present as ice throughout most of the disk. This is consistent with models of chemical processing during protostellar collapse by e.g. \citet{Visser09}, which suggests that most water ice is supplied to the disk from the protostellar envelope unaltered \citep[see also:][]{Yoneda16}. However, there are observations of crystalline ice in some disks, which only forms when water ice is heated close to its sublimation point, or first sublimes and then reforms \citep{McClure15}, which shows that also the record of external protostellar disks require a combination of chemical reset and inheritance.  This conclusion is supported by observations revealing thermal processing of small silicate particles in disks \citep{Waelkens96, Malfait98, vanBoekel04, watson09}.

Beyond CO, there are few observations of the chemical inventory in the protostellar disk interior. By contrast there is a growing number of observations showing interesting chemical structures at the edges and surfaces of protostellar disks. Figure~\ref{fig:PSD} shows that disk surfaces may present the same kind of complex organics traditional associated with hot corinos ({\it cf.} \citet{Bergner19b}). This discovery may be related to other observations of sharp chemical gradients from a carbon-rich, as traced by small hydrocarbons and CS, to an oxygen-rich chemistry, as traced by H$_2$CO and SO, at the centrifugal barrier where the gravitational pull of the nascent star is balanced by the centrifugal force \citep{Sakai14b,Sakai14a}. At this radius, infalling material is expected to experience a shock, which may result in partial sublimation of icy interstellar grains \citep{Aota15,Miura17}.
This ice sublimation may increase the gas-phase O/C ratio since ice mantles are O-rich, pushing the chemistry towards forming O-containing molecules at the expense of hydrocarbon destruction.
A difference between C-rich envelopes and SO-rich (O-rich?) protostellar disks are also seen in surveys of protostellar disks \citep[e.g.,][]{delaVillarmois18}, suggesting that chemical gradients across the centrifugal radius may be universal. They do not appear to always be associated with a rich organic chemistry, however, which suggests that such a chemistry is either a transient phenomenon, or sensitive to the abundance of organic feed-stock molecules inherited from the prestellar core stage.

\subsection{Summary: Chemistry in Protostellar Envelopes and Disks}

The protostellar stage is characterized by a combination of preservation of the major interstellar volatile C, N, and O carriers, by a transformation of a portion of simple interstellar organics into more complex ones, and by a partial chemical reset during the earliest stages of disk formation. As a result, predicting the disk composition at the outset of planet formation is complicated and far from a solved problem. All protoplanetary disks will likely be marked by a combination of interstellar, protostellar and disk chemical processes, but the relative importance of the three may vary between different disks dependent on the details of the protostellar collapse and disk formation \citep{Aikawa12, Drozdovskaya16}. More in depth studies of protostellar disks, further model exploration on the combined effects of chemistry and dynamics, as well as a better understanding of the volatile content of the young solar system are all key to map out the `typical' initial conditions of planet formation, and how they depend on the protostellar disk formation dynamics.

\section{Protoplanetary disk chemistry and planet formation}

In the final stage of forming a planetary system, the now pre-main sequence star has dispersed its envelope, and is only surrounded by a disk of gas and dust. There are several lines of evidence  that planet formation is well underway in these disks and they are therefore commonly referred to as protoplanetary disks or planet-forming disks \citep{Williams11, Andrews18,Andrews20}. The chemical structures of protoplanetary disks impact several aspects of planet formation. The time dependent radial and vertical distribution of molecules determines what material is available to planets forming at different disk locations and at different times. The division of molecules, and therefore volatile elements, between gas and solids constrains what material is available for incorporation into rocky planet cores versus primary atmospheres. The interplay between snowlines, grain compositions, and grain fragmentation and coagulation properties regulate pebble size distributions across the disk and therefore planetesimal and planet growth rates. Finally the locations of water and organic reservoirs in disks constrain how easy or difficult it is for young terrestrial planets to access these reservoirs. These different aspects of the chemistry of planet formation are reviewed in the following sub-sections. 

\begin{figure}[htbp]
\begin{center}
\includegraphics[width=0.8\textwidth]{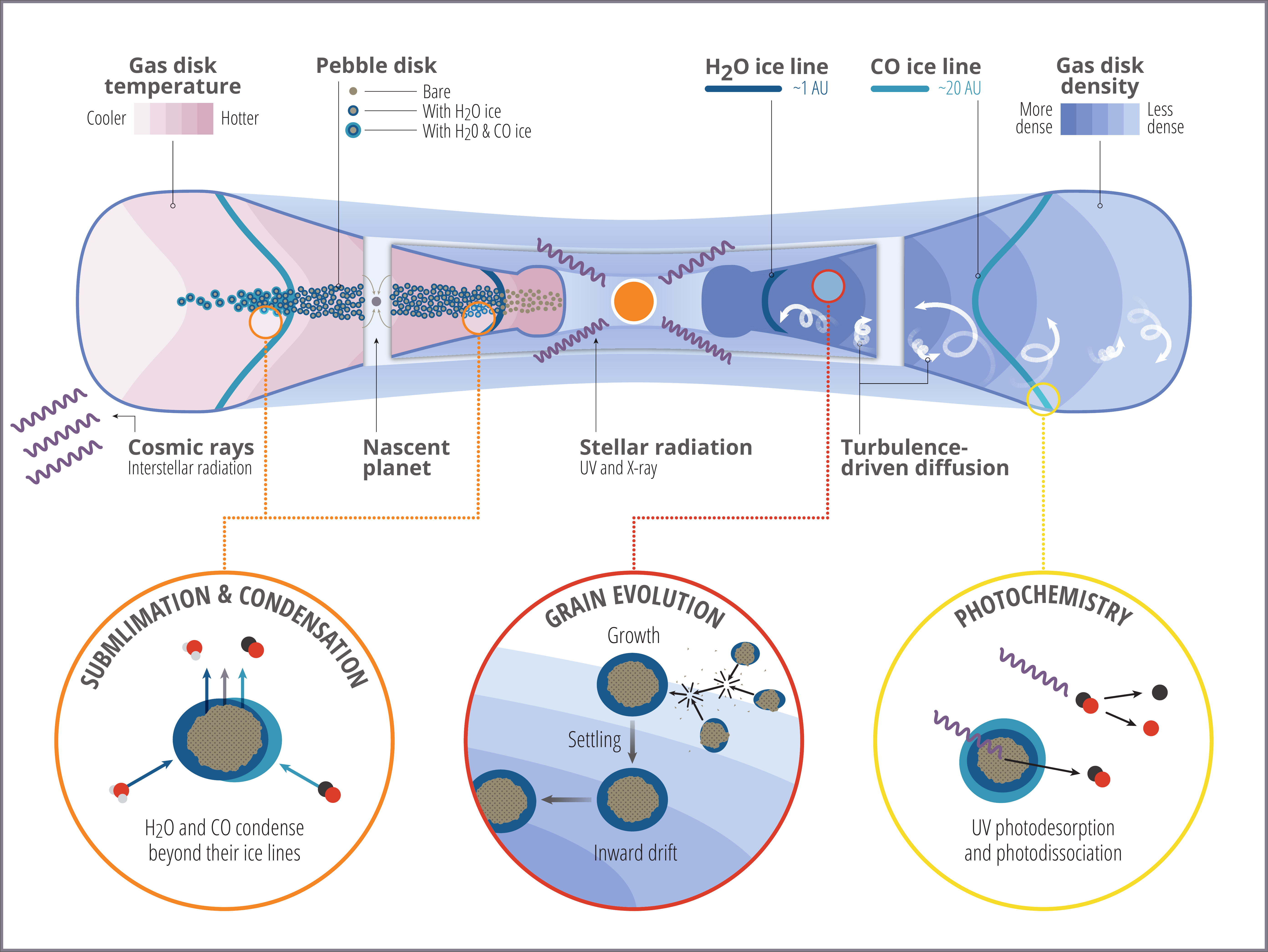}
\caption{Cartoon of protoplanetary disk chemistry and its coupling to different gas and grain dynamical processes.  Note the temperature gradient inward and upward (pink shading) resulting in a series of snowlines, as well as the inward and downward density gradient (blue shading). Disk surfaces are characterized by photon processes and display PDR-like chemistry. Disk midplanes are by contrast cold and UV-poor, and the main volatile reservoirs (other than H$_2$ and He) are in icy grain and pebble mantles. }
\label{fig:disk-cartoon}
\end{center}
\end{figure}

Before addressing the interplay between disk chemistry and planet formation, it is useful to review our current understanding of chemical and dynamical processes in protoplanetary disks as illustrated in Fig. \ref{fig:disk-cartoon}. There has been several reviews in recent years that have treated protoplanetary disks and their chemistry. we refer the reader to \citet{Henning13} for a review of disk chemistry and especially disk chemistry theory but there are also other numerous salient contributions \citep[e.g.][as a small sample]{Aikawa99, Gorti04, Nomura05, Woitke09, Kamp10, Walsh10, Bruderer12, Wakelam16, Schwarz19}.  Absent dynamical redistribution of material, the disk radial and vertical chemical structure is mainly set by gradients in temperature and ionizing radiation away from the central star, and down towards the midplane. This is analogous to clouds, and different disk regions can be profitably compared with interstellar structures that we have already discussed. The disk atmosphere resembles a PDR, while the disk midplane chemistry has many similarities with pre-stellar cores. The analogies are not perfect, however, because dynamical redistribution is not, in fact, absent, and because grain growth result in thermal and dynamical decoupling of dust and gas \citep{dAlessio01, Jonkheid04, Aikawa06, Nomura06, Nomura07}. Once grains grow they settle towards the midplane and drift inwards \citep{Blum08}, which can result in substantial redistribution of volatiles both vertically and radially \citep{Cuzzi04, Ciesla06, Meijerink09, Krijt16, Krijt18}.  Turbulent diffusion may also result in large-scale volatile transport. 

A major question is how efficient the chemical-dynamical evolution of disks is at resetting the chemistry, or whether chemical products inherited from previous stages of star and planet formation survive. The answer is complicated. Crudely, chemical time scales decrease inwards and upwards in disks, and we therefore expect inheritance to play the major role in the outer disk midplane, {\it in situ} chemistry in disk atmospheres, and a mixed behavior in intermediate regions. Dynamical transport of material from one region to another may change this picture dramatically, however, and we are only beginning to explore the effects of such transport on disk chemical structures during the epoch of planet formation \citep[e.g.][]{Semenov11}. 

With these caveats in mind, we here review aspects of disk chemistry of especial interest to predicting planet formation outcomes.
We first discuss how the disk chemical  structure can affect the planet formation process due to changes in grain growth and grain concentrations across volatile condensation/sublimation fronts or snowlines. Second we explore how snowlines combine with {\it in situ} disk chemistry and disk dynamics to set the distribution of volatile elements C, N and O during planet formation. Third we describe the distribution of organics during planet formation, and its dependence on inheritance and disk chemistry. Finally we review solar system constraints on the distribution of volatile elements during planet formation, and how water and organics can be incorporated into nascent, temperate and rocky planets, i.e. the kind of planets likely most hospitable to life.

\subsection{Snow lines and planet formation}

The dichotomy between dry terrestrial planets in the inner solar system, and water- and volatile-rich gas/ice giants and comets in the outer solar system has long be ascribed to a transition of water vapor to the solid state across the solar nebular snowline. There are several potential mechanisms through which snowlines regulate planet formation, some which apply only to the water snowline, and some which would also be active at the snowlines of other major volatiles such as CO, CO$_2$, CH$_4$, NH$_3$, and N$_2$. At the most basic level, snowlines always influence planet formation, since they determine the elemental content of solids and gas at different disk locations. This is reviewed in the next section. In this section we instead focus on whether and through which mechanisms snowlines can alter the efficiency at which planets form, resulting in preferred planet locations, and in predictable distributions of planet sizes. 

There are at least four mechanisms through which snowlines can affect the planet formation process by speeding up the growth of pebbles, the building blocks of planets and planet cores. First, across the snowlines the solid material surface density increases. This effect will be the largest for the most abundant volatiles. Exact numbers are unclear, but some estimates increases the column density of condensates (silicate rocks + ices) by  $\gtrsim$ 3 compared to right inside the snowline \citep{Hayashi81}. This alone might mean that solids grow to larger sizes beyond the water and CO snow lines. 
 
Second, diffusive flows of vapor across snowlines may both increase the column density of solids exterior to the snowline and result in preferential growth of larger particles.  \citet{Stevenson88} suggested that diffusive flow of water vapor outwards would result in condensation and accumulation of water ices via a so-called ``cold trap'' or ``cold finger'' effect. This effect can be greatly enhanced by the presence of radial drift of particles across the snowline \citep{Ciesla06,Ros13}:
in the presence of a radial pressure gradient the  gas orbits at sub-Keplerian velocities for a given radius \citep{Whipple72, Weidenschilling77}, resulting in a drag force on the solids, which induces radial drift of intermediate-sized (mm--m) grains and pebbles towards the pressure maximum. 
Diffusive flows are also expected in the vertical direction, where vertical temperature gradients produce snow surfaces that extend across the disk \citep{Willacy06, Meijerink09, Semenov11}. Exterior to the volatile snowline, vapor would diffuse downwards and be caught in the overall grain evolution (growth and settling) \citep{vanDishoeck14, Blevins16, Ros19}, further enhancing the midplane ice density \citep{Krijt16}.  Lower than expected water and CO vapor abundances beyond their respective snowlines support the presence of efficient vertical volatile flows towards the midplane \citep{Chapillon08, Hogerheijde11, Favre13, Krijt16, Krijt18}. This mechanism should operate along all major snowlines.

Third the material properties changes across snowlines, which may increase (or decease) stickiness and fragmentation thresholds, both of which aid (or quench) transforming initially sub-micron-sized particles into mm/cm-sized pebbles via coagulation \citep{Guttler10,Pinilla17}. Both theory  \citep{Dominik97, Wada13} and experiments \citep{Gundlach15} find that the velocity at which a grain-grain collision causes fragmentation, as opposed to growth, is an order of magnitude {\em higher} for water-ice coated grains as opposed to bare silicates.  Thus water ice coated grains are expected to readily grow to pebble sizes beyond the water snowline, while growth may be limited to sub-cm sizes in the innermost disk regions \citep{Birnstiel10}.  This break in dust size should be observable at mm-wavelengths because of the link between dust size population and the shape of the dust spectral emission as a function of wavelength \citep{Banzatti15}. Such a break has been detected in one object, V380~Ori,  which is undergoing a burst in its luminosity, likely tracing a temporary water snowline in the outer disk, which is resolvable by ALMA, due to an accretion event \citep{Cieza16}.
Further out in the disk where volatiles like CO$_2$ and CO freezes, the grain material properties change again, but this time perhaps destructively, resulting in diminished growth as a snowline is crossed \citep{Musiolik16,Pinilla17}.  This mechanism thus depends strongly on the kind of snowline.
 
Fourth, major snowlines may operate as  localized regions of enhanced pressure that reverses the overall pressure gradient in a disk.  The local pressure gradient will operate as a trap to drifting dust leading to a localized dust density enhancement and enhanced dust growth \citep{Cuzzi04}.  

\begin{figure}[htbp]
\begin{center}
\includegraphics[width=0.7\textwidth]{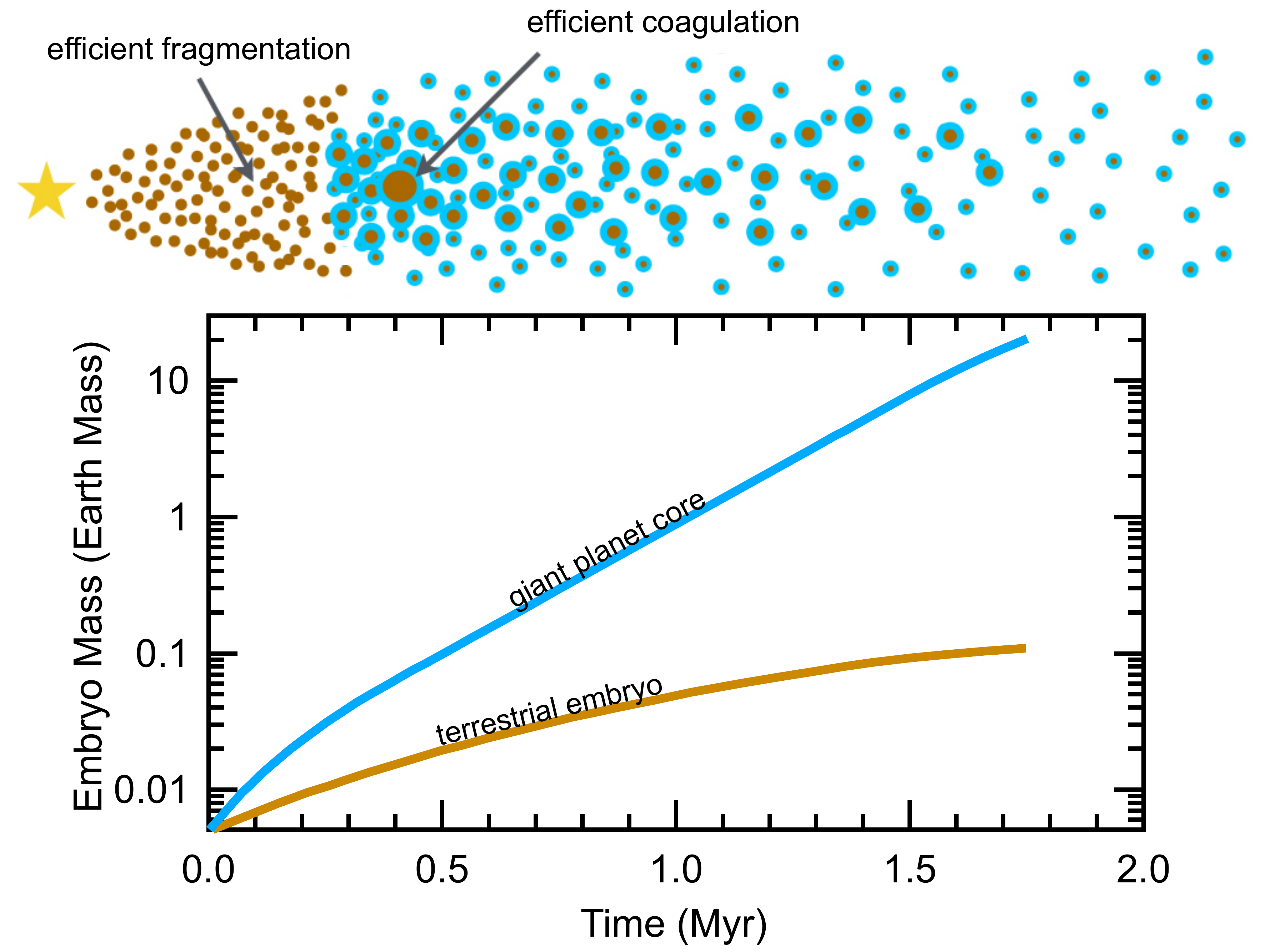}
\caption{({\em Top:}) Schematic of grain growth across the water snowline according to experimental results and theory.  Schematic taken from \citet{Banzatti15}.
({\em Bottom:}) Mass growth due to pebble flux for two simulations: one inside the water snowline, shown as a brown line, and another outside the water snowline, given as the blue line.  Figure taken from \citet{Morbidelli15}.  Based on the work of \citet{Lambrechts14} the 20 Earth mass embryo will capture nebular gas and become a giant planet.
\label{fig:SnowLineImplication}}
\end{center}
\end{figure}
 
What might all this mean for planet formation near the snowline of the most abundant volatiles?  
\citet{Morbidelli15} explored planetary growth via pebble accretion \citep[e.g.,][]{Johansen17} assuming that pebbles are few cm-sized beyond the water snowline and mm-sized interior to the snowline.  The resulting growth of planetary bodies is shown in Fig.~\ref{fig:SnowLineImplication}.  Beyond the ice line the larger mass of cm-sized pebbles allows for growth of a giant planet core within a few Myr, while only a Mars-sized embryo forms inside the snow line. Several other models also find a change in planet formation at the water snowline \citep{Armitage16, Schoonenberg17, Drazkowska18}, which together with a recent observational coincidence between the outer edge of the H$_2$O emission region and the 13~au dark lane identified by ALMA in HL Tau \citep{ALMA15}, suggest that the water snowline is important for regulating planet growth in disks  \citep{Salyk19}
 
 A central question is whether we should only expect preferential dust growth and planet formation at the water snowline, or also at the snowlines of  other major volatiles, such as CO, CO$_2$ or N$_2$, as posited e.g. by \citet{Cuzzi04}.   \citet{Zhang13} suggested that this might be the case based on a comparison of one realization of the midplane temperature profile and observations of significant symmetric sub-structure (rings/gaps) within the dust emission towards HL Tau \citep[see also][]{Pinilla17}.   They argued that this substructure was indicative of a chance in dust properties due to localized grain growth at snowlines.
 More recent comparisons of dust sub-structure in many disks do not find a link of substructure with the inferred sublimation temperature of key volatiles \citep{Huang18, Long18}.     At face value, this suggests that the snowlines of volatiles such as CO, CO$_2$, N$_2$, etc are not inducing significant dust evolution.  However, for most disks we still lack direct observations of snowline locations (see below), and laboratory data suggest that they may appear at quite different temperatures in different disks because the ice sublimation temperatures depend sensitively on ice composition and morphology \citep{Fayolle16}.

\begin{figure}[htbp]
\begin{center}
\includegraphics[width=0.7\textwidth]{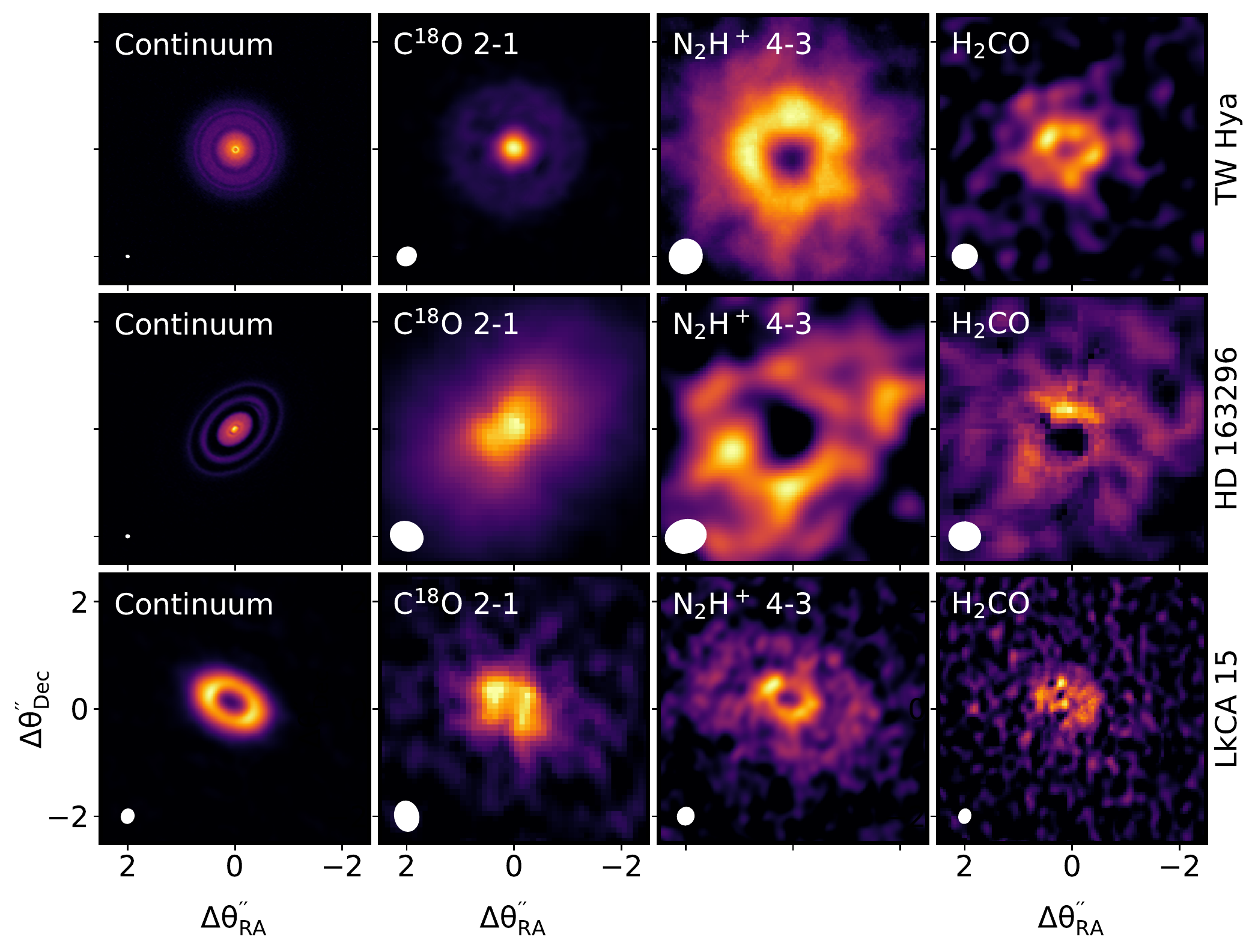}
\caption{ALMA gallery of observed CO snowline tracers towards three disks, shown together with their millimeter continua images (left panel). {\it 2nd panel:} C$^{18}$O emission is expected to be optically thin in outer disk regions and provides constraints on radial CO depletion. {\it 3rd panel:} N$_2$H$^+$ is readily destroyed by CO gas and should  trace midplane regions where CO is frozen out, i.e. the inner rim of the N$_2$H$^+$ ring should coincide with the CO snowline. {\it 4th panel:} H$_2$CO forms through CO ice hydrogenation, among other pathways, and should be present in excess exterior to the CO snowline. The synthesized beams are shown in the lower left corner of each panel. Data are taken from \citep{Huang18,Andrews18,Zhang17,Pegues20,Qi13c,Qi19,Oberg17}}
\label{fig:obs-snowlines}
\end{center}
\end{figure}

 Direct observations of the water snowline is hard because water sublimates close to the star limiting our ability to spatially resolve emission \citep{Notsu16, Carr18, Notsu19}.  Instead, the best constraints exist for CO. The CO snowline is easier to observe than the other snowlines because it occurs at large distances, which makes it easier to resolve, and because there are multiple potential molecular tracers of CO midplane freeze-out: optically thin CO isotopologues, H$_2$CO and N$_2$H$^+$. Figure \ref{fig:obs-snowlines} shows examples of snowline tracers towards three disks, TW Hya, HD 163296 and LkCa 15. Direct observations of CO isotopologues is the conceptually most straightforward method of detecting CO snowlines; at the location of the CO midplane snowline there should be a sharp decrease in CO emission. This appear to be the case for TW Hya and other systems \citep{Zhang17, Qi19, Zhang19}, but it can be difficult to distinguish between a drop in CO and overall gas column density. 
 N$_2$H$^+$ emission should trace CO snowlines because it is only observable in disk regions where CO is depleted from the gas phase \citep{Qi13c}, while H$_2$CO should trace CO freeze-out because it readily forms from hydrogenation of CO ice \citep{Oberg17}. The two chemical snowline tracers can also be difficult to interpret, however, because their identifications with CO freeze-out are not unique; N$_2$H$^+$ can also trace CO-poor disk atmospheres \citep{vantHoff17}, and H$_2$CO can also form through gas-phase chemistry that is independent of CO freeze-out. The community is still working out for which kind of disk which method provides the most accurate CO snowline radius estimate \citep[e.g.][]{Zhang19,Qi19}, and how to extend the lessons learned from CO to other molecules \citep{Qi19}.

\subsection{Disk volatile structures and budgets}
\label{sec:diskandplanetcomp}

The distributions and abundances of abundant volatile elements in the disk regulates the bulk compositions of gas giants, as well as the volatile inventories of terrestrial worlds \citep[e.g.,][]{Oberg11e,Bergin15}. In a static disk the elemental distribution is fundamentally set by a combination of  chemistry, which determines the main carriers of O, C, N, S, and P, and the freeze-out temperatures of these carriers. The compositions of inner disk volatiles has been reviewed by \citet{Pontoppidan14}, and we focus here on the outer disk, i.e. the disk exterior to the water snowline. Based on models and comparisons between cloud and cometary abundances \citep{Cleeves13,Drozdovskaya19}, the volatile composition exterior to the water snowline in disks is to a large extent inherited from the molecular cloud, and interstellar cloud chemistry may therefore be more important than disk chemistry to understand the main reservoirs of volatiles in disks. 
Given the chemical composition outlined in Fig. \ref{fig:pie-chart}, molecular condensation/freeze-out temperatures then determines where in the disk the elements are in solid form and available for incorporation into planetesimals and planet cores, and where it is in gas and can only become incorporated into planets through nebular gas accretion. 

\begin{figure}[htbp]
\begin{center}
\includegraphics[width=0.3\textwidth]{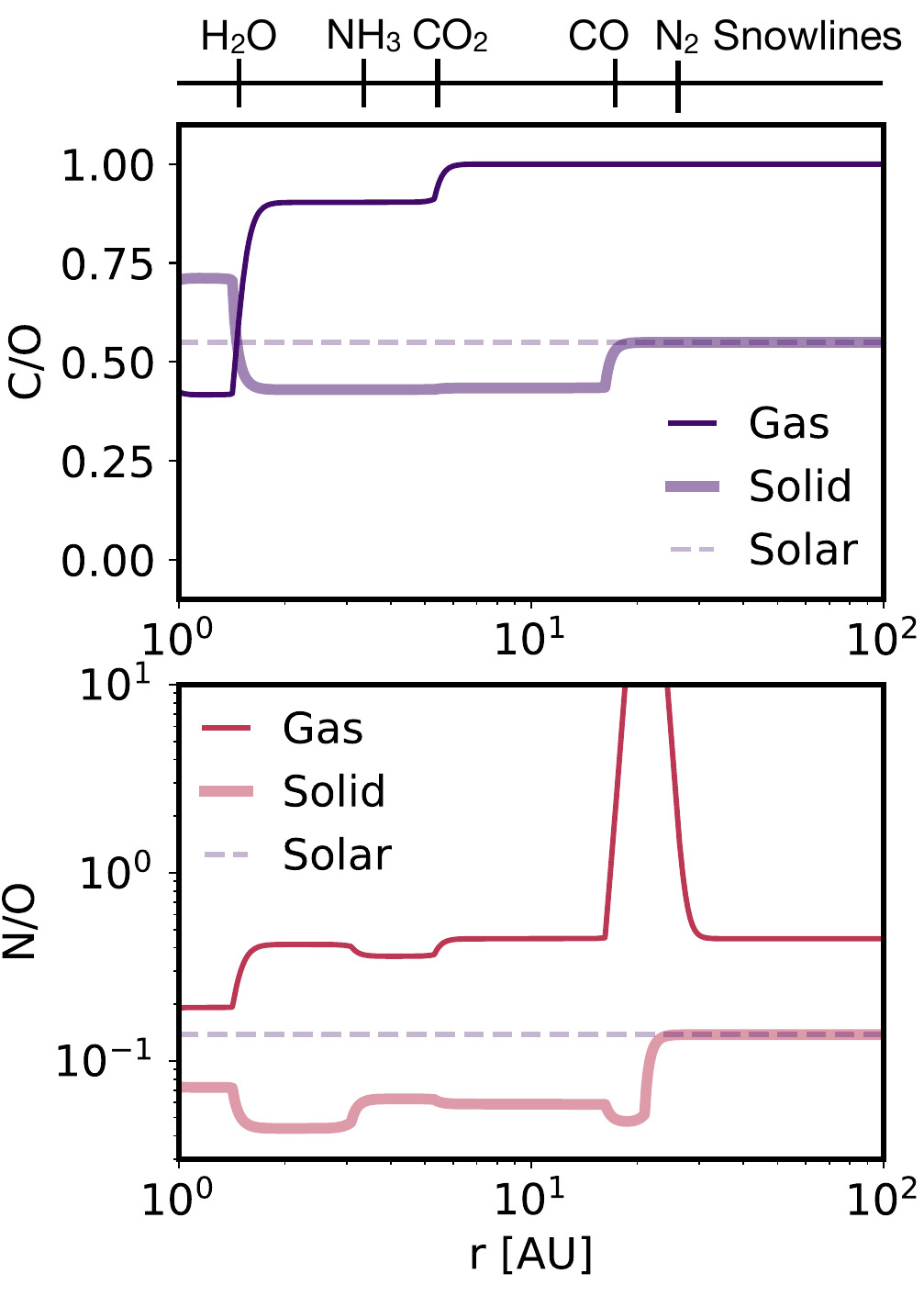}
\caption{The sequential freeze-out of volatile carriers of O, C and N in protoplanetary disks results in radially dependent C/O and N/O ratios in gas and solids. For this model we used the distribution of O, C and N carriers shown in Fig. \ref{fig:pie-chart}, assuming that the unknown O is equally divided between water and refractories, and that the unknown N is in N$_2$. Beyond the CO and N$_2$ snowline we assume that a small amount of CO and N$_2$ (10$^{-3}$) is maintained in the gas through non-thermal desorption. }
\label{fig:snowlines}
\end{center}
\end{figure}

Figure \ref{fig:snowlines} shows that using a model of the Solar Nebula \citep{Oberg19}, and assuming that the division of O, C and N between different volatile carriers established in the molecular cloud stage survives disk formation, the gas and solid state C/N/O ratios change across the disk because of different sublimation temperatures of major carriers of O, C and N. In particular, the difference in sublimation temperature of water (a major O carrier), and the hypervolatile  CO and N$_2$ (major C and N carriers, respectively) cause the gas-phase C/O and N/O ratios to increase with disk radius \citep{Oberg11e,Piso16}. As a result a Gas Giant forming exterior to the water snowline should initially have a O-rich core and a O-poor atmosphere, though subsequent mixing between the phases may change this picture \citep{Helled18, Oberg19}. Whether such planets are common will be tested by upcoming exoplanet missions, most notable by observations by JWST.

There are several dynamical processes that are active in disks that complicate the simple scenario outlined above, with static, narrow snowlines regulating the distribution of volatiles. First, snowline locations are fundamentally set by the disk thermal profile, which is regulated by stellar radiation and accretion, both which decrease over the lifetime of the disk. Simultaneously the disk mass decreases, however, increasing the efficiency at which it can be heated. The net result is a complex dependence of snowline locations on disk age \citep{Dodson-Robinson09}. Second, sublimation time scales are often similar to grain drift timescales due to angular momentum transfer between Keplerian solids and sub-Keplerian gas. This can result in that snowlines are effectively moved inwards by as much as 50\% of their `static' radius \citep{Ciesla06,Piso15}. Third, diffusive flows from interior of the snowline to the volatile poor exterior can deplete the gas in regions substantially inward of the snowline, while not changing the location where ice formation begins \citep{Ros13}. 
 Pebble drift can also redistribute volatiles from the inner to the outer disk \citep{Cuzzi04,Oberg16b,Krijt18}. Finally all these dynamical processes occur together with a chemical evolution that can change the nature of the disk volatiles, e.g. converting substantial amounts of CO into CO$_2$ and/or CH$_3$OH. This does not {\it per se} change snowline locations, but rather it changes which snowlines matter for setting the C/N/O ratios in solids and gas across the disk \citep{Meijerink09, Reboussin15, Schwarz16, Schwarz18, Bosman18, Eistrup18}. 

These complications aside, the expected preferential sequestration of O and, to some extent, C into solids in  outer disk regions should have observable effects on the gas phase chemical composition, which should become increasingly O and C poor. This has been verified by cold water observations by Herschel towards two nearby disks, DM Tau and TW Hya \citep{Bergin10,Hogerheijde11}, supplemented by a survey \citep{Du17}. However, to everybody's surprise water was not just depleted in the disk midplane (as expected), but also high up into the disk atmosphere where UV radiation from the central star should be able to maintain some of the water in the gas-phase if it had been present at normal abundances in icy grain mantles.   The inferred water ice depletion in disk atmospheres from far-IR observations is roughly consistent with observed water ice features at infrared wavelengths \citep{Debes13, Honda16}.   More recent observations have revealed an underabundance of CO in disk atmospheres indicative of that carbon is also more depleted from the gas than expected from passive freeze-out alone \citep{Favre13,Schwarz16,Miotello16,Cleeves18}. By contrast nitrogen does not appear to be depleted in disk atmospheres \citep{Cleeves18, Anderson19}, which likely reflects both the volatile nature of the main nitrogen carrier, N$_2$, and the difficulty of converting N$_2$ into more refractory species on disk time scales. There is mixed evidence for C-depletion interior to the CO snowline \citep{Salyk11,Bosman17,Zhang17,Bosman19,Zhang19}, and mapping out the C/O and N/O ratios at at scales of 10s of au and less is an observational frontier \citep{Najita13}. 

Several possible removal mechanisms have been proposed to explain the missing O and C in disk atmospheres, including grain settling, diffusive flows, and chemical conversion of CO into more refractory species \citep[e.g.][]{Reboussin15, Krijt16,Yu17, Eistrup18, Schwarz18, Bosman18}. The latter explanation may only be possible to explore by observing the volatile composition inside the water snowline where all volatiles, including potential CO derivatives, such CH$_3$OH should return to the gas.  \citet{Meijerink09}, \citet{Salyk11}, and \citet{Blevins16} used data from Spitzer, and in some cases Herschel \citep[see also summary in][]{Pontoppidan14} to model water vapor emission at radii commensurate with its snowline.  These models are consistent with water returning to the gas-phase at $\sim$ISM abundance levels inside the water snow line. However, there is some uncertainty because water emission is optically thick, JWST is needed both to establish the water abundance, and the composition of any co-sublimating organic ice.

\subsection{Disk organic chemistry}

The distributions of organic molecules in disks are of special interest because of their connection to complex organic molecules in comets and asteroids \citep{Ehrenfreund00,Altwegg19}, and therefore possible connection to prebiotic chemistry on planets following asteroid and comet bombardment \citep{Pearce17,Howard13}. A range of organic molecules have been observed in protoplanetary disks: HCN, CN, C$_2$H$_2$, C$_2$H, $c$-CH$_3$H$_2$, H$_2$CO, CH$_3$OH, HCOOH, HC$_3$N, H$_3$CN, CS and H$_2$CS. Most of these have been detected at millimeter and sub-millimeter wavelengths \citep{Dutrey97,Aikawa03,Thi04,Chapillon12,Oberg15,Walsh16,Favre18,LeGal19}, but HCN is also observed at IR wavelengths, and C$_2$H$_2$ is only observed at these shorter wavelengths due to its lack of a permanent dipole moment \citep{Carr08,Salyk08, Najita10}. In most cases IR and mm observations probe upper disk layers, and the connection between observed organic inventories to those of planet-forming material in disk midplanes is not straightforward. There is both chemical and kinematic evidence for a dynamical link between midplane and disk atmosphere chemical abundances, however. First, observations of complex organic molecules such as CH$_3$CN are only reproduced in models where cold icy grains are lifted up into the disk atmosphere \citep{Oberg15}. Second, \citet{Teague19} recently showed that there are meridional flows in at least one disk, which would effectively circulate disk material between the atmosphere and midplane. While we currently lack models that can quantify the importance of this link for the organic composition of forming planetesimals, a first step is a detailed understanding of the organic inventory where we can observe it, and this is reviewed below. 

Constraints on the organic reservoirs in inner disk regions, commensurate with formation zones of habitable planets comes from IR observations; observations at longer wavelengths are confined to larger disk radii with existing facilities. 
Such IR observations of HCN and C$_2$H$_2$ have revealed that small organic molecules are abundant in the inner disk atmosphere \citep{Pontoppidan14}. There is a curious difference in composition between disks around Solar-type and cool stars -- the latter generally lack HCN \citep{Pascucci09}. There are several possible explanations for this difference, with different implications for the organic content of planet-forming pebbles and boulders around different kinds of stars. If HCN is produced {\it in situ} in the inner disk gas, the observed differences in HCN/C$_2$H$_2$ ratio may simply be due to different radiation fields around Solar-type and cool stars, and may not be informative about the organic composition of planet-forming organic solids in the two kinds of disks. While this explanation remains a possibility it is not supported by current astrochemical disk models \citep{Walsh15}. 

A second possible explanation is that the observed chemistry depends on disk dynamics, which depends on disk mass, which in its turn correlates with stellar mass \citep{Andrews13}.  \citet{Najita13} found that  the HCN-water ratio depends on disk mass, which is explained by regulation of the C/O ratio by drift, or the lack thereof, of water-rich pebbles into the inner disk, which in its turn regulates the HCN gas-phase production. A third possible explanation for the observed differences in HCN/C$_2$H$_2$ between disks around cool and Solar-type stars is that this reflects a difference in organic composition of inward drifting pebbles, i.e. that the organic chemistry in the outer disk depends on the stellar mass and luminosity.
Supporting this scenario is the similarity between HCN/C$_2$H$_2$/H$_2$O abundance ratios in disks around Solar-type stars and solar system comets \citep{Pontoppidan14,Altwegg19}. This would imply that planets form in chemically very different environments around different kinds of stars, which may drive substantially different prebiotic chemistries.  

\begin{figure}[htbp]
\begin{center}
\includegraphics[width=0.6\textwidth]{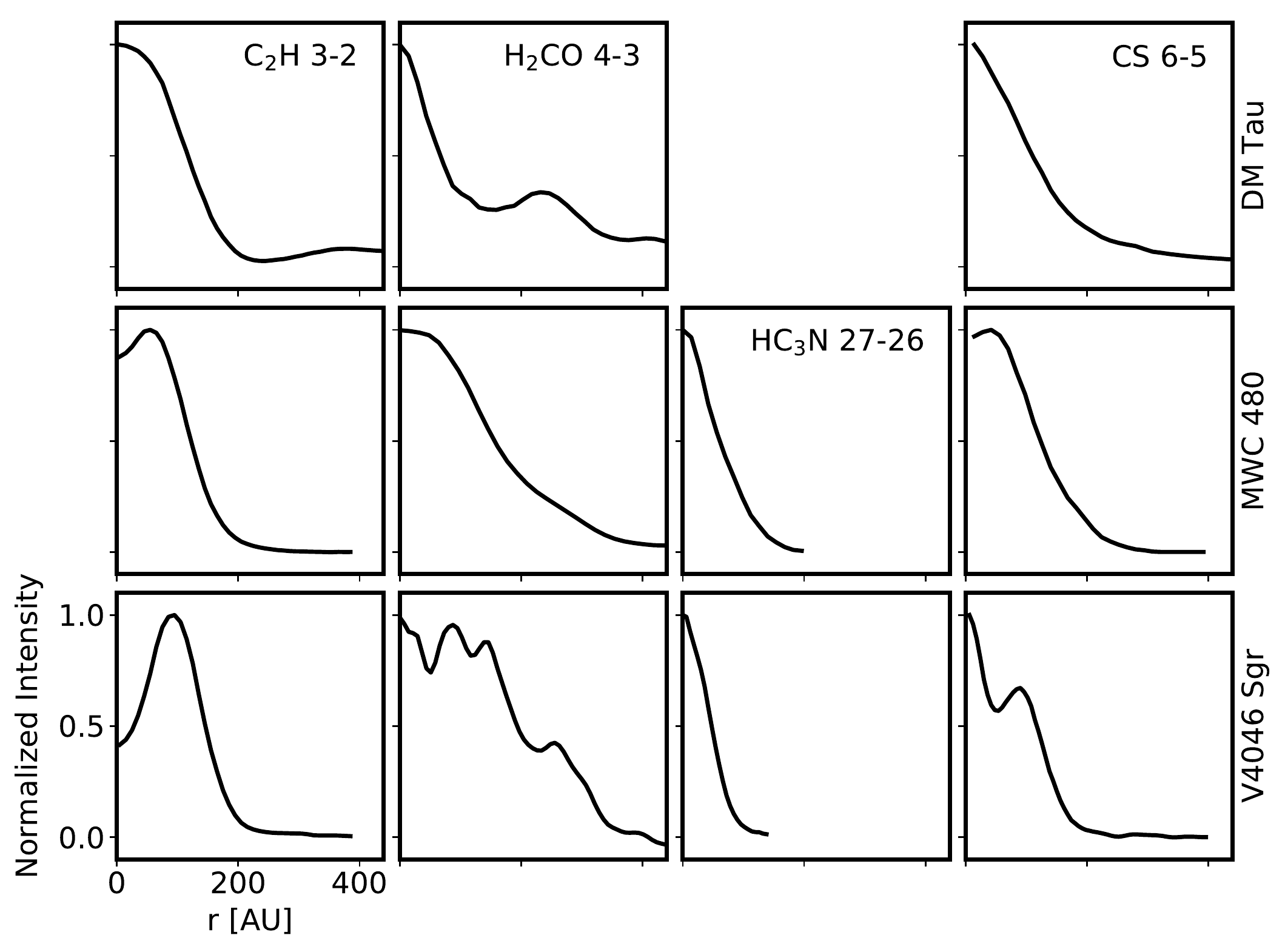}
\caption{Radial profiles of four commonly observed organic molecules towards three representative protoplanetary disks: the young T Tauri disk DM Tau, the Herbig Ae disk MWC 480, and the old T Tauri disk V4046 Sgr. The four molecules, C$_2$H, H$_2$CO, HC$_3$N, and CS, are examples of the hydrocarbons, O-, N-, and S-containing organics.  Note that the distribution of organic molecules differ both between disks, and between different kinds of organics.
The origins of most variations are currently highly speculative, but are thought to be connected to differences in C and O depletions across and between disks, snowline locations, and dust sub-structure. Data from  \citep{Bergner19b,LeGal19, Pegues20}.  }
\label{fig:disk-org}
\end{center}
\end{figure}

Spectrally resolved millimeter and sub-millimeter observations probe disk chemical compositions at larger disk radii, typically beyond 10s of au. Spatially resolved observations of organic molecules exist towards samples of $\sim$10 disks, and Fig. \ref{fig:disk-org} shows radial profiles of four organic molecules towards a subset of these disks. The molecules have been chosen as representatives of the four families of organic molecules observed in disks: hydrocarbons (C$_2$H), O-containing organics (H$_2$CO), nitriles and iso-nitriles (HC$_3$N), and S-containing organics (CS). Based on these profiles alone, (1) organics are not evenly distributed across disks, (2) there are multiple kinds of organic distributions and by extension organic chemical pathways present in each disk (compare e.g. the distribution of C$_2$H and H$_2$CO in V4046 Sgr), and (3) a single molecule can be distributed differently in different disks. This entails that if the observed chemical distributions trace or influence planetesimal compositions, the planetesimal organic compositions will strongly depend both where and around which star the planetesimals formed. 

A number of disk chemical models have been developed to predict and explain the kind of observations introduced above. \citet{Henning13} provides a quite recent review of these models. Some patterns that are emerging is that it is difficult to explain the relative abundances of nitrogen and oxygen bearing organics unless the disk atmospheres are C-rich and/or O-poor \citep{Bergin16, Kama16, Cleeves18,LeGal19b}, that O-bearing organics are present at lower abundances than expected \citep{Walsh16}, and that grain-surface chemistry is needed to reproduce both O and N-bearing complex organic abundances \citep{Walsh16,Loomis18}.

\subsection{Elemental Accounting in the Solar System}

Much of the existing connective tissue between astronomical composition and exoplanet composition links measurements of C/H and O/H (C/O) in planetary atmospheres, such as summarized by \citet{Madhusudhan19}, to disk observations and planet formation theory.  Fig.~\ref{fig:snowlines} is a prime motivator for this work.    In our solar system, Jupiter and Saturn appear to be enriched in both C and N with values above that of the proto-Sun \citep{Wong04, Atreya20}.  This has been posited to be the result of volatile enrichment in the gas via pebble drift \citep{Cuzzi04, Oberg16b} but see also \citet{Helled18}.  The N enrichment for Jupiter has been suggested to be the result of formation of Jupiter beyond the N$_2$ iceline followed by migration \citep{Oberg19, Bosman19-n2}.  The uncertain disposition of oxygen in this context is critical as the C/O ratio is a useful link to the formative conditions, but oxygen is difficult to probe in gas and ice giants \citep[e.g.,][]{Cavalie20}. In Jupiter, Juno measured O/H $= 2.7^{+2.4}_{-1.7}$ times protosolar \citep{Li20}, potentially consistent with enrichment, but this needs to be measured at different latitudes to determine if this represents the bulk composition. 
 
\begin{figure}[htbp]
\begin{center}
\includegraphics[width=1.0\textwidth]{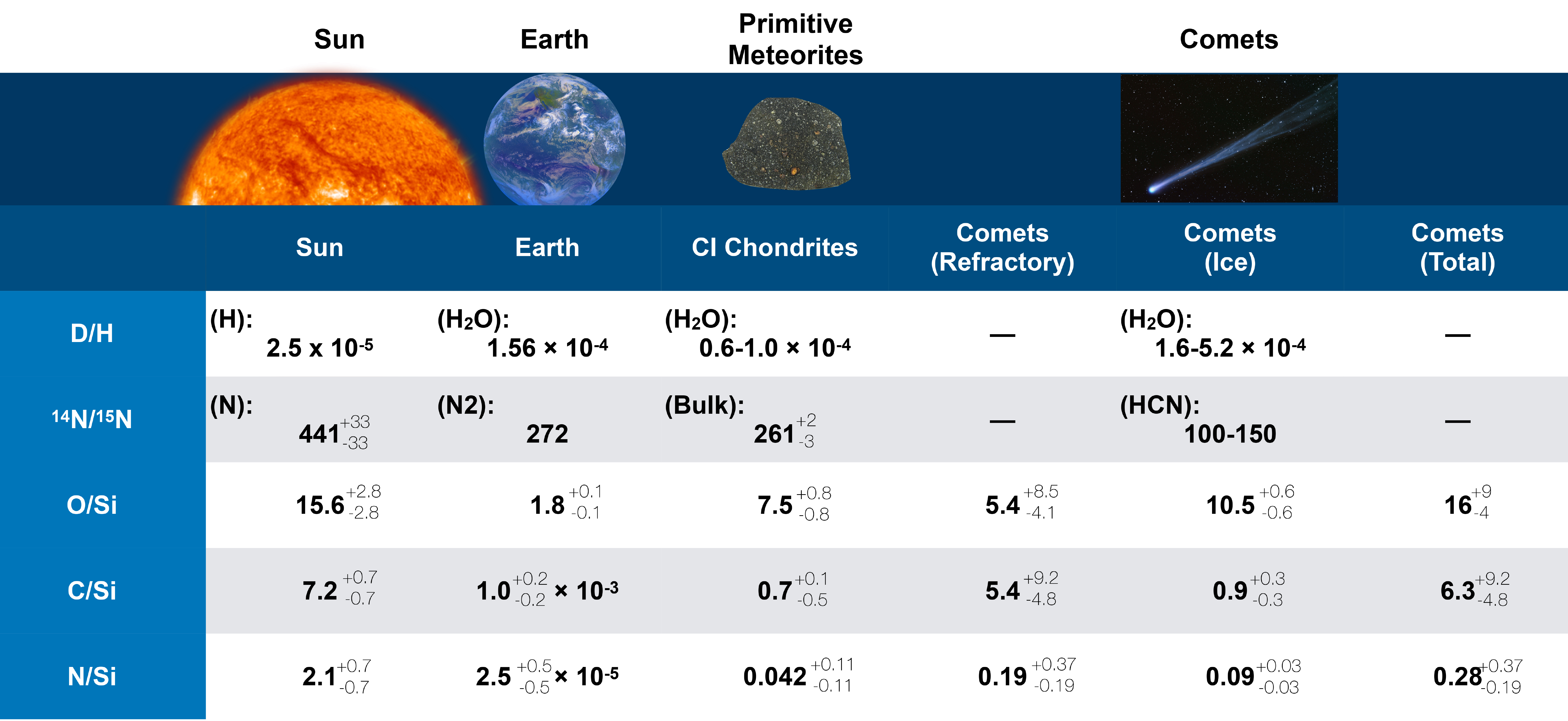}
\caption{Volatile element budget and isotopic composition in select Solar System objects.
(Sun): Image credit SeekPNG.  Protostolar D/H ratio as determined by  \citet{Geiss72} and nitrogen isotopic ratio of the solar wind from Genesis \citep{Marty11}.  Atomic elemental abundance ratio of O, C, and N to Si are solar system abundances from 4.56 Gyr ago as estimated by \citet{Lodders10}.  (Earth): Image credit NOAA.  D/H ratio of Earth water is set by Vienna Standard Mean Ocean Water. Atomic ratios are taken from \citet{Wang18} for O and from \citet{Bergin15} for C and N. (CI Chondrites:) Image  Isotopic and atomic ratios for CI chondrites as given by \citet{Wanke88} and \citet{Wang19}. Errors represent the range in the samples. D/H ratio for water are from \citet{Alexander12}.  (Comet:) Image of Comet ISON by Waldemar Skorupa.  Comet water D/H and HCN nitrogen isotopic ratios represent the range detected in both Oort Cloud and Jupiter family comets \citep{Mumma11, Lis19}.   Bulk elemental abundance atomic ratios are from Comet 67P/Churyumov-Gerasimenko as provided by \citet{Rubin19} for a dust/ice ratio of unity. }
\label{tab:budgets}
\end{center}
\end{figure}

Elemental accounting within the rocky/icy bodies are also of use and 
in Fig. \ref{tab:budgets} we provide an accounting of the budgets of the volatile elements O, C, and N in key solar system objects dominated by rocks and ices.   We continue our focus C, O, and N as they are the most abundant and well-characterized of the volatile elements, but see below for a brief discussion of sulfur and phosphorous.
We do not mention H, which is carried terrestrial worlds via H$_2$O and giant planets via H$_2$.    There are some distinctions in terms of the accounting in Fig. \ref{tab:budgets} that are worth mentioning. For isotopic ratios, we note which species we are referring to, e.g. H$_2$ or H$_2$O.   
For comets we provide the accounting within volatile ices and the total content which includes refractory material and the ices. Finally, CI Chondrites refer to a class of meteorites that are the least altered; hence they represent most primitive sampling of rocky material available \citep{Weisberg06}.

Inside the water snowline, the fate of the C and N that was initial locked up into refractory material is a puzzle. For oxygen it is certain that any O locked in silicates in the ISM finds its way into solar system rocks, since silicates are only destroyed very close to the star where temperatures rise above $\sim$1500~K \citep{Dullemond10}, and this is consistent with the compositions of Earth and carbonaceous chondrites.  But, materials provided to the disk also had large content of C and N carried in refractory form, which does not appear to have survived.
 The Earth acquired as little as 1 in 10,000 carbon atoms available at formation \citep[Table~\ref{tab:budgets}][]{Finocchi97, Lee10, Pontoppidan14} and CI chondrites, which formed in the inner Solar Nebula, are significantly carbon-depleted \citep{Geiss72,Bergin15, Rubin19}.  By contrast, substantial refractory and volatile carbon is found in comets, which perhaps best reflect the initial inventory of the solar system.
 Nitrogen exhibits even more extreme depletions in inner solar system solids, and is also under-abundant in comets (consistent with N$_2$ being a major nitrogen carrier), though comets still retain some more refractory nitrogen content \citep{Altwegg20}.  
 
 It is important to state that constraints on the Earth are somewhat uncertain due to possible sequestration of a significant amount of carbon (or nitrogen) deep within the Earth \citep{Dasgupta13}.  Regardless, it is clear that there is a gradient in the carbon and nitrogen content of solids from the inner to outer solar system; both carbon and nitrogen must have been primarily in volatile forms during early formative stages of pre-cursor materials inside of, at least, $\sim 3$~au.
This may be a generic result in planet formation as the remnants of other rocky planetary systems currently accreting onto white dwarfs are also carbon-poor \citep{Jura15, Xu19}.   Various theories have been proposed to account for this disposition \citep{Gail02, Lee10, Bergin15, Anderson17, Gail17, Klarmann18}.

   Sulphur and Phosphorus are two other abundant volatile elements of import to biology.  For sulphur, there is a puzzle in the interstellar medium as the main carrier remains unidentified  \citep[][and references therein]{Kama19}. In the solar system, the accounting is clearer: C I chondrites contain a nearly solar abundance of S \citep{Lodders10}, and so does the comet 67P/Churyumov–Gerasimenko \citet{Calmonte16}.  \citet{Kama19} used photospheric abundances of B stars, which have radiative envelopes with longer mixing timescales, to infer that in the accreting disk material, $\sim 89\pm8$\% of sulphur resides in refractory material. In the case of 67P, suflur is present in refractory, semi-refractory and volatile (H$_2$S) comet phases, however \citet{Calmonte16}, and it is possible that the volatility of sulfur reservoir changes between the inner and outer disk regions.
   
 Similar to sulphur, chondritic and solar composition are in close agreement for P/Si, indicative of that most phosphorous was refractory in the inner Solar Nebula \citep{Lodders10, Wang19}. Also similar to sulfur, there is potential tension between this finding and the identification of a substantial volatile phosphorous reservoir in comets. Most notably, phosphorous is detected in the coma of comet 67P, indicative of that at least some phosphorous is present in more volatile form during planet formation \citep{Altwegg19}. \citet{Rivilla20} identified the volatile phosphorous in 67P with PO, and found that the P/O abundance was consistent with solar, indicative of a low fraction of refractory P in the comet-forming zone, in direct contrast with the conclusion drawn from meteoritic abundances. It is currently unclear whether there is a gradient in refractory and volatile phosphorous carriers across disks, or whether 67P presents an unusual phosphorous abundance pattern. The solar system measurements can be compared with interstellar observations: 
PO and PN have been detected in the vicinity of protostars, most often in shocked outflow regions   \citep{Caux11, Yamaguchi11, Rivilla16, Lefloch16,Rivilla20,Bergner19d}. Their abundances are well below solar phosphorous abundances, indicative of that most phosphorous in the dense ISM is solid and hidden from view. This could be consistent with either the meteoritic or cometary inferences, and more data is needed to identify the nature, and especially the volatilty, of phosphorous and sulfur carriers during the early stages of star and planet formation.

\subsection{Delivery of water and organics to temperate planets}

The supply of water and organic material to Earth  and other temperate terrestrial worlds during their formative stages, is a central question for habitability.  How do key ingredients of life that originate in interstellar space affect the surface compositions of worlds that lie within the habitable zone, and make them chemically habitable.  Based on the discussion in the earlier sections it is clear that the Earth, and many other rocky worlds, formed within their respective snowlines.  This does not mean that these planets formed dry, as our own planets composition tells us; rather the supply of both water and organic material is likely part of the formation process.   Below we will describe the theories that have been proposed to account for the origin of Earth's water and, by extension, its organics, and how lessons learned from Earth can be extended to other habitable planets. 

When elucidating this process we will especially  consider  the D/H and $^{14}$N/$^{15}$N isotopic ratios of Earth's volatiles.   The D/H ratio hints at Earth's water history via a simple fact. Earth's water, and indeed all solar system water \citep{Cleeves14}, has an excess of deuterium relative to hydrogen above that within the main deuterium repository HD.   The same is also the case for interstellar water.
As discussed above deuterium fractionation reactions for water proceed provided the gas temperature is less than 30~K. The opposite is also true: any deuterium enrichment, formed at cold temperatures and released to the gas via sublimation, will be redistributed back into HD on a timescale of few $\times$10$^5$~yr or less depending on the strength of radial and/or vertical motions \citep{Drouart99, Furuya13}.    The expected temperature of Earth's building blocks was clearly in excess of the sublimation temperature of water ice ($\sim$150--200~K).  At these temperatures the deuterium signature would be destroyed and the resulting water would not be D-enriched.  Thus, Earth's water received a contribution in full or in part from water ice beyond the snowline where the cold temperatures kept water frozen in the ices and the deuterium enrichment intact.   

Nitrogen isotopic enrichments constitute another tool that has been used to explore the origin of Earth's volatile content \citep{Marty12}.  Tracing the origin of this enrichment to particular locations in the ISM or in the solar nebula is difficult as there are multiple pathways to $^{15}$N enrichments, through low temperature reactions and N$_2$ isotopic self-shielding, which could have been active at multiple stages during the pre-Solar, proto-Solar and Nebular stages \citep{Wirstrom18, Furuya18, Visser18}.  Still, $^{14}$N/$^{15}$N similarities and differences between the Earth and different solar system volatile reservoirs can be used to constrain the origin of the Earth's nitrogen within the solar system (Table~1), though
more work needs to be done to understand the origin of the $^{15}$N excess \citep[e.g.,][]{Hily-Blant19}.

There are (at least) three theories for the origin of water on a dry world --  in situ, inward motion of the water snowline, and dynamical supply from beyond the ice line --  and here we evaluate them in light of the Earth's isotopic enrichment patterns:  

\begin{enumerate}[leftmargin=0.5cm]
    \item {\em In Situ:}  In general, it is thought that interstellar molecules are bound to grain surfaces via physical adsorption and, for water, via H-bonds.   However, interstellar grains are likely complex with more fractal like surfaces as seen in laboratory experiments of coagulation \citep{Blum08}.  These surfaces themselves may present sites where water might be chemisorbed onto dust grains or grouped into more resistant clusters, leading to a dilute water presence on dust grains even above the nominal evaporation temperature \citep{Stimpfl06, Muralidharan08, King10}. The water vapor needed for this model is typically assumed to originate from a complete reset of the chemistry in the hot ($>$2000~K) cooling nebula model that is suggested for numerous inner solar system solids  \citep[see, e.g.,][]{McDonough95, Lodders10}.  \citet{Dangelo19} explored this question and demonstrated that hydrated minerals (phyllosilicates) could form and provide 0.5-10 Earth oceans worth of water to forming terrestrial worlds even at 300 K.  This would essentially mean that the Earth formed wet.   At face value this solution would provide water, but it is not certain whether carbon/nitrogen would also be provided at the same time.  We note, that absent a high-temperature D-enrichment pathway such as suggested by \citet{Thi10}, the water would carry no deuterium enrichment in this scenario, which would require additional, substantial delivery of cold material, 
    to match the Earth's D/H ratio.  
      
     \item {\em Snowline evolution:}  During the early stages of nebular evolution the midplane is viscously heated via accretion,  keeping the nebular snowline beyond 1 AU.  However, the accretion rate decays with age \citep{Hartmann16}, and when it falls below $\sim$10$^{-8}$~M$_\odot$/yr the snowline is predicted to move interior to 1 AU \citep{Garaurd07, Oka11}.   At this time ($\sim 1$~Myr) the disk is clearly dominated by mm/cm sized pebbles \citep{Andrews18}.  \citet{Sato16} and \citet{Ida19} show that if sufficiently large planetary embryos ($\sim0.1$M$_{\oplus}$) exist at 1 AU after the snowline moves inwards, then these embryos can grow via accretion of ice-coated pebble as the pebbles drift inwards.  In addition to water, these pebbles would contain some carbon, and nitrogen, and likely carry the isotopic signatures of chondritic material.  Thus, this scenario could be consistent with the isotopic evidence.  However, \citet{Morbidelli16} suggested that if Jupiter were to form  before the water snowline moved interior to 1 AU, it would effectively stop icy-pebble drift leaving the inner nebula mostly dry.  For systems with giant planets present, such as our own, this scenario may then not provide an explanation of Earth's enrichment in water and other volatiles. By contrast, in systems that lack Jupiter-analogs at $3-7$~au, and they are estimated to constitute $>$90\% of planetary systems  \citep{Wittenmeyer20}, water and organics will likely be present  in high abundance during early stage of terrestrial planet formation \citep{Raymond20}.
     
     \item {\em Dynamical Supply:}  This is the traditional model for water delivery to habitable planets in systems similar to our own.
     In the solar system, there is a close connection between planetesimal bombardment of terrestrial planets and the presence of Jupiter, and its interaction with the gaseous and planetesimal disk.  It is currently unclear whether planetary systems that lack a Jupiter analog could use this scenario to deliver volatiles to habitable worlds.
    In the context of the Earth, the appeal of this scenario, is that provides a natural supply of water and organics from the outer asteroid belt, and match isotopic constraints.   There are two comprehensive reviews in the recent literature that explore these models within the context of water supply and the reader is referred to these references \citep{Obrien18, Raymond20}.  
     \end{enumerate}
     
     Of course, these different theories need not be exclusive and it is possible that combinations contribute.  Further,  here we only discuss the initial supply for water and C/N/S/P bearing organic/silicate material.  In subsequent planet-evolutionary stages, there are a number of potential loss mechanisms that would modify the original volatile inventory.  A brief (and not exhaustive) list of such mechanisms include $^{26}$Al heating dehydration \citep{Lichtenberg19}, impact-driven atmospheric loss \citep{Schlicting15}, thermal metamorphism \citep{Hashizume98}, and core formation \citep{Bergin15, Grewal19}.  In all, the assembly of habitable planets starts with basic physical and chemical principles that drive the formation of abundant water ice and organic material in interstellar space, but the implementation of these principles is complex and a research frontier.
     
\subsection{Disk Chemistry Summary}

Planets form in disks around pre-main sequence stars and the volatile composition of disks therefore directly impacts the volatile composition of planets. Over the past decade direct observations of disk gas have provided important clues about the disk volatile inventory, revealing e.g. that outer disk regions are oxygen depleted in the gas, and therefore, presumably, oxygen-enriched in its solids. This points to an active disk chemistry and a changing volatile composition during planet formation. These changes cannot be too large, however, since the similarity between cometary and interstellar ice compositions, as well as the water deuterium record in the solar system point to substantial inheritance of interstellar volatiles. Figuring out the exact balance between inheritance and disk chemistry is a key question for the next decade.

Within disks the distribution of volatiles may affect how efficiently planet formation proceeds. Snowlines are thought to change the grain growth process, through differential sticking and fragmentation, diffusive flows, and induced pressure traps. Snowlines also change the division of elements between gas and solids, and therefore regulate the initial composition of cores and atmospheres on planets, including their elemental C/N/O ratios.   In this regard, bulk composition is a key tool to link planet formation to both gas giants and terrestrial worlds.

In addition to bulk compositions, we have paid special attention to the distribution of water and organics across disks, and the likelihood that they may become incorporated into or delivered to terrestrial planets. Disk do present a rich organic chemistry, and likely inherit much of the complex organic chemistry that is observed in protostellar envelopes. These potential precursors of prebiotic molecules are widely distributed across disks, but there is growing evidence that the organic composition varies both across and between different disks, indicating that planets may form with a diverse set of initial organic chemical conditions. How water and organic molecules do become part of the surface layers of planets, and populate their atmospheres and hydrospheres is not yet a solved problem. In the solar system impact delivery appears to have played an important role, but how important is debated. It is also unclear how easy it is to transfer this mechanism of enriching terrestrial planets in water and organics to exoplanetary systems without an analog to our outer solar system.
     
\section{Review Summary and Future Outlook}

In this contribution we have explored how the chemistry of interstellar space influences the composition of planetary systems.   Our focus has been on the elemental pools of the most abundant volatile elements C, O, N and H, on water, and on  organics.  

The chemical trajectory begins during the transition from diffuse to dense clouds, when the ultraviolet (UV) radiation exposure of all cloud material decreases, but the decrease is more dramatic towards the cloud center, resulting in temporal and spatial chemical gradients. During this process, the chemistry relevant for future planet formation shifts from the gas-phase, which regulates CO and N$_2$ formation, to the grain surface, where water and feedstock organics form. Importantly, at this stage most volatile carbon become locked into CO , most N into N$_2$, and most volatile O into CO and H$_2$O.  The majority of these volatiles appears to survive the star formation process and become incorporated into planet-forming disks intact. The bulk volatile composition is thus set long before planet formation begins. {\it A key mystery at this stage is that one third of the oxygen is unaccounted for, which affects e.g. C/O ratio predictions throughout the protostellar and planet-forming stages.}
 
Organics represent a small portion of the overall elemental inventory, but their proposed connection to prebiotic chemistry motivates a detailed understanding of their formation and transformations during star and planet formation. The initial organic feed-stock molecules are formed in cold cloud regions through gas-phase chemistry (unsaturated organics) and grain-surface processes (saturated organics). During cloud collapse, heating by protostars and shock-induced grain heating, increases the mobility of ice constituents, resulting in a rich and complex organic chemistry. This chemistry is qualitatively quite well understood, but a lack of quantitative experimental and computational data on chemical reaction probabilities and rates limits the predictive capability of existing models. {\it Ice experiments, comprehensive surveys of protostellar chemistry, and theoretical models that fully couple the dynamics and time-dependent chemistry constitute frontiers in protostellar organic chemistry.}

Models predict that most of the bulk volatile composition, as well as the organic chemistry developed in the protostellar stage is incorporated into the planet-forming disk. Within the disk some alteration is expected, especially in the inner disk, and at disk surfaces. In the latter regions, there is clear evidence for substantial oxygen depletion, some carbon depletion, and no nitrogen depletion, which must affect the elemental make-up of pebbles and planetesimals in the planet-forming disk midplane. The balance between inheritance and chemical processing is currently being explored theoretically, through comparisons between solar system and interstellar/protostellar observations, and through direct observations of disks with ALMA and other millimeter and sub-millimeter facilities. A difficulty in establishing this balance is that it certainly depends on a coupling between time-dependent chemistry and disk dynamics, especially mixing between disk surfaces and midplanes, and radial volatile transport through diffusive flows, accretion flows, and pebble drift. {\it There are several ongoing efforts to connect chemical theory and dynamics, and we expect rapid model improvement in the next few years. These models will be tested by forthcoming observations of disk gas by ALMA, and with JWST observations of midplane ices, and inner disk surface gas.}

The final step is how to connect chemical compositions of gas and pebbles in disks with the compositions of young planets. This requires a better understanding of disk midplane compositions (see above), but also on how volatiles can be added to planets post-formation. The latter is especially important to predict the water and organic content of temperate, Earth-like planets. {\it It is currently not clear under which conditions Earth-like planets can sample volatiles formed or preserved in outer disk regions, beyond the water snowline.} More in-depth cometary studies are key to assess their formation zones, as well as their relationship to terrestrial volatiles.  A frontier in the connection between astrochemistry and planet composition regards the carbon content of inner disks; in the ISM, 50\% of carbon resides in  refractories, which appears preserved in comets, but seems to have been lost in the inner solar system. Depending on the nature of the refractory carbon removal mechanism, terrestrial planets may be generally carbon-poor, and depend on impacts both for water and organic delivery.

Finally, we note that Astrochemistry is an inherently interdisciplinary field. Its past and future successes depend on a combination of astronomical observations, chemical physics laboratory experiments, quantum calculations, molecular dynamics theory, and astrochemical models.   We are  entering an exciting era where astrochemistry is connecting with planetary and exoplanetary science to explore the formation of planets and the evolution of their hydrospheres and atmospheres. While the chemistry of planet formation sets the initial conditions of planets, the atmospheric chemistry and geochemistry determines how these initial conditions develop, and how often we may expect the complex chemistry we believe preceded life here on Earth.

\bigskip
\noindent EAB acknowledges support from NSF AAG Grant (\#1907653) and the following grants from NASA's Exoplanetary Research (80NSSC20K0259) and Emerging Worlds (80NSSC20K0333) programs. KI\"O acknowledges support from the Simons Foundation (SCOL \#321183) and an NSF AAG Grant (\#1907653). The authors are grateful to Viviana Guzmán, Romane Le Gal, Jennifer Bergner, Jamila Pegues, Chunhua Qi,  and Chin-Fei Lee for contributing to the figures. The authors are also deeply appreciative to 
Henrick Beuther,
Geoff Blake,
Paola Caselli,
Cecilia Ceccarelli,
Ilse Cleeves,
Rachel Freisen,
Robin Garrod,
Maryvonne Gerin,
Viviana Guzmán,
Javier Goicoechea,
Eric Herbst,
Pierre Hily-Blant,
Romane Le Gal,
Klaus Pontoppidan,
Evelyne Roueff,
Nami Sakai,
Ian Sims,
Vianney Taquet,
Ewine van Dishoeck,
Qizhou Zhang
for providing input on the content of the review.





\bibliographystyle{elsarticle-harv}



\end{document}